\documentclass[twocolumn,pra,showpacs,aps]{revtex4}

\usepackage{amsmath}
\usepackage{amsbsy}
\usepackage{amsthm}
\usepackage{amssymb}
\usepackage{latexsym}
\usepackage[dvips]{color,graphicx}

\begin{document}

\title{Ground-state properties of hard-core bosons confined on \\
one-dimensional optical lattices}

\author{Marcos Rigol}
\affiliation{Institut f\"ur Theoretische Physik III, Universit\"at Stuttgart, 
Pfaffenwaldring 57, D-70550 Stuttgart, Germany.}
\author{Alejandro Muramatsu}
\affiliation{Institut f\"ur Theoretische Physik III, Universit\"at Stuttgart, 
Pfaffenwaldring 57, D-70550 Stuttgart, Germany.}

\begin{abstract}
We study the ground-state properties of hard-core bosons trapped by 
arbitrary confining potentials on one-dimensional optical lattices. 
A recently developed exact approach based on the Jordan-Wigner 
transformation is used. We analyze the large distance behavior of the 
one-particle density matrix, the momentum distribution function, and 
the lowest natural orbitals. In addition, the low-density limit in the 
lattice is studied systematically, and the results obtained compared 
with the ones known for the hard-core boson gas without the lattice.
\end{abstract}
\maketitle

\section{Introduction}

\vspace{-0.19cm}
Low dimensional systems have been the subject of increasing interest
over the past decades, traditionally in condensed-matter physics, 
and recently in the framework of quantum gases. There, recent advances 
in atom waveguide technology \cite{thyw99,muller99,dekker00,key00,bongs01}, 
the realization of quantum gases in very anisotropic traps \cite{schreck01,
gorlitz01} and loading Bose-Einstein condensates (BEC) on optical
lattices \cite{greiner01,moritz03,stoferle03}, allowed experimentalists 
to obtain a rich variety of systems where the reduced dimensionality 
rules the physics. Particularly interesting is the case in which the
quantum dynamics of the system becomes quasi-one-dimensional. In
that case it was shown \cite{olshanii98,petrov00,dunjko01} that in 
certain regimes of large scattering length, 
low densities, and low temperatures, bosons behave as 
impenetrable particles known as hard-core bosons (HCB's).

The one-dimensional (1D) homogeneous gas of HCB's was 
introduced by Girardeau \cite{girardeau60}, and a one-to-one
correspondence between 1D HCB's and spinless fermions was
established. It was shown later by Lenard and by Vaidya and Tracy
\cite{lenard64,vaidya79} that in homogeneous space, 
the 1D HCB gas does not exhibit true condensation at zero temperature 
since there is only off-diagonal quasi-long-range order 
(one-particle correlations decay as a power law), 
leading to an occupation of the lowest effective single-particle 
state (the highest occupied one) $\sim \sqrt{N_{b}}$, 
where $N_{b}$ is the total number of HCB's. 

Recently, theoretical studies have been focused on the ground-state 
properties of 1D HCB gases in the presence of harmonic potentials 
\cite{girardeau01,minguzzi02,lapeyre02,papenbrock03,forrester03,gangardt04},
which are required in the experiments with quantum gases to keep
the particles confined. Using the Fermi-Bose mapping, quantities like 
density profiles, momentum profiles, the natural orbitals (effective 
single-particle states), and the one-particle density matrix have been 
calculated for a finite number of particles. Some results were 
extended to the thermodynamic limit ($N_b \rightarrow \infty$) 
\cite{forrester03,gangardt04}. It has been found that similarly to the 
homogeneous case the occupation of the lowest natural orbital is 
$\sim\sqrt{N_b}$ \cite{papenbrock03,forrester03}, so that only a 
quasicondensate develops in the trap.

Another case of interest is given by HCB's on a lattice. 
These systems have been realized experimentally very recently 
\cite{paredes04}. In the periodic case, the 1D HCB Hamiltonian can be 
mapped into the 1D $XY$ model of Lieb, Schulz, and Mattis \cite{lieb61}.
This model has been extensively studied in the literature. 
At zero temperature the asymptotic behavior of the correlation functions is 
known \cite{mccoy68,vaidya78,mccoy83}. 
In the presence of a confining potential, the systems realized 
experimentally in Ref.\ \cite{paredes04}, the situation is more 
complicated. This is because the trapping potential is 
equivalent to the addition of a space-varying transverse field to the $XY$ 
Hamiltonian, for which in general analytical results are not available.

In this work we study ground-state properties of
1D trapped HCB's in a lattice. The lattice 
opens new possibilities for engineering states that cannot be obtained 
in continuous systems. In particular, for HCB's it is possible to create 
pure Fock states when the occupation in some regions of the system reaches 
the value $n=1$, such that coherence is lost there, and these sites decouple 
from the rest of the system. In addition, the properties of the systems 
without the lattice can be recovered in the low-density limit in a lattice. 
We follow a numerical approach, based on the Jordan-Wigner
transformation \cite{jordan28}, that allows us to calculate exactly 
the one-particle Green's function and analyze arbitrary confining potentials.  
This approach was recently introduced in Ref.\ \cite{rigol04_1} 
to study ground-state properties, and generalized in 
Ref.\ \cite{rigol04_2} for the study of the nonequilibrium dynamics. 

We concentrate here on the behavior of the off-diagonal part of the 
one-particle density matrix, the momentum distribution function, and the 
natural orbitals. These quantities have a different behavior for HCB's when 
compared with spinless fermions to which the HCB Hamiltonian is mapped 
by means of the Jordan-Wigner transformation. We find that the one-particle 
density matrix $(\rho_{ij})$ decays as a power law $\sim |x_i -x_j|^{-1/2}$ 
for large distances, like in the periodic case \cite{mccoy68}, 
irrespective of the confining potential chosen. This is valid even when 
portions of the system reach occupation $n_i=1$, such that coherence is lost 
there. The power law above is shown to determine the scaling of the 
occupation of the lowest natural orbital in the thermodynamic limit. 
This scaling and its finite-size corrections are also studied for arbitrary 
powers of the confining potential. In addition, the low-density limit in the 
lattice is analyzed in detail and the results obtained are compared 
with the ones presented in Ref.\ \cite{forrester03} for continuous systems.

The exposition is organized as follows. In Sec.\ II we describe
the numerical approach used. In Sec.\ III we discuss the properties 
of HCB's on periodic lattices. Systems confined in harmonic traps are 
analyzed in Sec.\ IV. In Sec.\ V we generalize to arbitrary powers of the 
confining potential the results obtained for the harmonic case. 
Finally, the conclusions are given in Sec.\ VI.

\section{Exact approach}

\vspace{-0.19cm}
In the present section we describe in detail the exact approach we 
follow to study 1D HCB's in the presence of a lattice \cite{rigol04_1}. 
The HCB Hamiltonian can be written as 
\begin{equation}
\label{HamHCB} H = -t \sum_{i} \left( b^\dagger_{i} b^{}_{i+1}
+ h.c. \right) + V_\alpha \sum_{i} x_i^\alpha \ n_{i },
\end{equation}
with the addition of the on-site constraints
\begin{equation}
\label{ConstHCB} b^{\dagger 2}_{i}= b^2_{i}=0, \  
\left\lbrace  b_{i},b^{\dagger}_{i}\right\rbrace =1. 
\end{equation}
These constraints on the creation ($b^{\dagger}_{i}$) and 
annihilation ($b_{i}$) operators avoid double or higher occupancy.  
Notice that the brackets in Eq.\ (\ref{ConstHCB}) 
apply only to on-site anticommutation relations, while 
$[b_{i},b^{\dagger}_{j}]=0$ for $i\neq j$, i.e., they commute 
on different sites \cite{lieb61}. In Eq.\ (\ref{HamHCB}), 
the hopping parameter is denoted by $t$, and the last term  
describes an arbitrary confining potential, with power $\alpha$ and 
strength $V_{\alpha}$; $n_{i}= b^{\dagger}_{i}b_{i}$ is the 
particle number operator. 

In order to obtain the ground-state properties of HCB's, we 
use the Jordan-Wigner transformation \cite{jordan28},
\begin{equation}
\label{JordanWigner} b^{\dag}_i=f^{\dag}_i
\prod^{i-1}_{\beta=1}e^{-i\pi f^{\dag}_{\beta}f^{}_{\beta}},\ \ 
b_i=\prod^{i-1}_{\beta=1} e^{i\pi f^{\dag}_{\beta}f^{}_{\beta}}
f_i \ ,
\end{equation}
which maps the HCB Hamiltonian into the noninteracting spinless 
fermions Hamiltonian
\begin{eqnarray}
\label{HamFerm} H =-t \sum_{i} \left( f^\dagger_{i}
f^{}_{i+1} + h.c. \right)+ V_\alpha \sum_{i} x_i^\alpha \
n^f_{i },
\end{eqnarray}
where $f^\dagger_{i}$ and $f_{i}$ are the creation and
annihilation operators for spinless fermions and $ n^f_{i }=
f^\dagger_{i}f_{i}$ is their particle number operator. This 
means that HCB's and fermions have exactly the same spectrum. 
The nontrivial differences between both systems 
appear in the off-diagonal correlation functions as shown below.

Using the mapping above, the one-particle Green's function 
for the HCB's can be written in the form
\begin{eqnarray}
\label{green1} G_{ij}&=&\langle \Psi^{G}_{HCB}|
b^{}_{i}b^\dagger_{j}|\Psi^{G}_{HCB}\rangle \nonumber \\
&=&\langle \Psi^{G}_{F}| \prod^{i-1}_{\beta=1}
e^{i\pi f^{\dag}_{\beta}f^{}_{\beta}} f^{}_i f^{\dag}_j
\prod^{j-1}_{\gamma=1} e^{-i\pi f^{\dag}_{\gamma}f^{}_{\gamma}} 
|\Psi^{G}_{F}\rangle \nonumber \\
&=&\langle \Psi^{A}_{F}|\Psi^{B}_{F}\rangle,
\end{eqnarray}
where $|\Psi^{G}_{HCB}\rangle$ is the HCB ground-state wave function, 
and $|\Psi^{G}_{F}\rangle$ is the equivalent noninteracting fermion 
ground-state wave function. In addition, we denote
\begin{eqnarray}
\label{states} \langle \Psi^{A}_{F}|&=&\left( f^{\dag}_i
\prod^{i-1}_{\beta=1} e^{-i\pi f^{\dag}_{\beta}f_{\beta}}
|\Psi^{G}_{F}\rangle \right)^\dag, \nonumber \\
|\Psi^{B}_{F}\rangle &=&f^{\dag}_j \prod^{j-1}_{\gamma=1} e^{-i\pi
f^{\dag}_{\gamma}f_{\gamma}} |\Psi^{G}_{F}\rangle.
\end{eqnarray}

The ground-state wave function of the equivalent Fermionic system
can be obtained by diagonalizing Eq.\ (\ref{HamFerm}) 
(the properties of such systems were analyzed in 
Ref.\ \cite{rigol03_3}), and can be written in the form
\begin{equation}
\label{wavefunct} |\Psi^{G}_{F}\rangle=\prod^{N_f}_{\delta=1}\ 
\sum^N_{\sigma=1} \ P_{\sigma \delta}f^{\dag}_{\sigma}\ |0 \rangle,
\end{equation}
with $N_f$ the number of fermions ($N_f=N_b$), $N$ the number of 
lattice sites, and the matrix of the components ${\bf P}$ is given by 
the lowest $N_f$ eigenfunctions of the Hamiltonian,
\begin{equation}
{\bf P}=\left(
\begin{array}{c c c c c c}
P_{11}&P_{12}&\cdot&\cdot&\cdot&P_{1N_f} \\
P_{21}&P_{22}&\cdot&\cdot&\cdot&P_{2N_f} \\
\cdot&\cdot&\cdot&\cdot&\cdot&\cdot \\
\cdot&\cdot&\cdot&\cdot&\cdot&\cdot \\
\cdot&\cdot&\cdot&\cdot&\cdot&\cdot \\
P_{N1}&P_{N2}&\cdot&\cdot&\cdot&P_{NN_f} 
\end{array}
\right)
\end{equation} \\

In order to calculate $|\Psi^{A}_{F}\rangle$ (and 
$|\Psi^{B}_{F}\rangle$) we notice that
\begin{equation}
\prod^{i-1}_{\beta=1}e^{-i\pi f^{\dag}_{\beta} f_{\beta}}
=\prod^{i-1}_{\beta=1} \left[ 1-2\ f^{\dag}_{\beta}f_{\beta} \right].
\end{equation}
Then, the action of $\prod^{i-1}_{\beta=1} 
e^{-i\pi f^{\dag}_{\beta}f_{\beta}}$ on the Fermionic ground state 
[Eq.\ (\ref{wavefunct})] generates only a change of sign on the elements 
$P_{\sigma \delta}$ for $\sigma \leq i-1$, and the further creation 
of a particle at site $i$ implies the addition of one column to 
${\bf P}$ with the element $P_{iN_f+1}=1$ and all 
the others equal to zero (the same can be done for $|\Psi^{B}_{F}\rangle$). 
Then $|\Psi^{A}_{F}\rangle$ and $|\Psi^{B}_{F}\rangle$ can be written as
\begin{eqnarray}
\label{states1} | \Psi^{A}_{F}\rangle&=&\prod^{N'_f}_{\delta=1}\ 
\sum^N_{\sigma=1} \ P'^{A}_{\sigma \delta}f^{\dag}_{\sigma}\ |0 \rangle, 
\nonumber \\
|\Psi^{B}_{F}\rangle &=&\prod^{N'_f}_{\delta=1}\ 
\sum^N_{\sigma=1} \ P'^{B}_{\sigma \delta}f^{\dag}_{\sigma}\ |0 \rangle,
\end{eqnarray}
where ${\bf P'}^{A}$ and ${\bf P'}^{B}$ are obtained from ${\bf P}$ 
changing the required signs and adding the new column ($N'_f$=$N_f$+$1$). 

The Green's function $G_{ij}$ is then calculated as
\begin{eqnarray}
\label{determ}
\langle \Psi^{A}_{F}|\Psi^{B}_{F}\rangle
&=&\langle 0 | \prod^{N'_f}_{\delta=1}\ \sum^N_{\sigma=1} \ 
P'^{A}_{\sigma \delta}f_{\sigma} 
\prod^{N'_f}_{\bar{\delta}=1}\ \sum^N_{\bar{\sigma}=1} \ 
P'^{B}_{\bar{\sigma} \bar{\delta}}f^{\dag}_{\bar{\sigma}}\ 
|0 \rangle \nonumber \\ 
&=&\sum^N_{ \sigma_1,\cdot \cdot \cdot,\sigma_{N'_f} \atop
{\bar{\sigma}}_1,\cdot \cdot \cdot,\bar{\sigma}_{N'_f}}
P'^{A}_{\sigma_11}\cdot \cdot \cdot P'^{A}_{\sigma_{N'_f}N'_f}
P'^{B}_{\bar{\sigma}_11}\cdot \cdot \cdot P'^{A}_{\bar{\sigma}_{N'_f}N'_f}
\nonumber \\ &&\times
\langle 0 |f_{\sigma_1}\cdot \cdot \cdot f_{\sigma_{N'_f}}
f^{\dag}_{\bar{\sigma}_{N'_f}} \cdot \cdot \cdot f^{\dag}_{\bar{\sigma}_1}  
|0 \rangle \nonumber \\
&=&\det\left[ \left( {\bf P}^{'A}
\right)^{\dag}{\bf P}^{'B}\right],
\end{eqnarray}
where the following identity was used
\begin{equation}
\langle 0 |f_{\sigma_1}\cdot \cdot \cdot f_{\sigma_{N'_f}}
f^{\dag}_{\bar{\sigma}_{N'_f}} \cdot \cdot \cdot f^{\dag}_{\bar{\sigma}_1}|0 \rangle
=\epsilon^{\lambda_1\cdot \cdot \cdot \lambda_{N'_f}}
\delta_{\sigma_1\bar{\sigma}_{\lambda_1}}\cdot \cdot \cdot
\delta_{\sigma_{N'_f}\bar{\sigma}_{\lambda_{N'_f}}},
\end{equation}
with $\epsilon^{\lambda_1\cdot \cdot \cdot \lambda_{N'_f}}$ the 
Levi-Civita symbol in $N'_f$ dimensions, the indices $\lambda$ 
have values between one and $N'_f$. 

The Green's function (\ref{determ}) is then evaluated numerically, and 
the one-particle density matrix is determined by the expression 
\begin{equation}
\rho_{ij}=\left\langle b^\dagger_{i}b_{j}\right\rangle =G_{ij}
+\delta_{ij}\left(1-2 G_{ii} \right).
\end{equation}

An alternative approach to the one presented here to calculate the 
one-particle density matrix of HCB's confined on optical lattices, 
using the Jordan-Wigner transformation, 
was followed by Paredes {\it et~al.} \cite{paredes04}. 
In their work the elements of the one-particle density matrix were 
evaluated as T\"oplitz determinants of matrices with sizes up to 
$(N-1)\times (N-1)$. Within our approach only determinants of 
$(N_b+1)\times (N_b+1)$ matrices are evaluated. 
    
\section{Hard-core bosons on periodic systems}

In this section we analyze the properties of HCB's in 1D periodic 
lattices. In this case the Hamiltonian (\ref{HamHCB}) can be written as
\begin{equation}
\label{HamHCBHom} H = -t \sum_{i} \left( b^{\dag}_{i} b_{i+1}
+ h.c. \right).
\end{equation}
This Hamiltonian, with the additional constraints (\ref{ConstHCB}), 
is particle-hole symmetric under the transformation 
$h_i=b^{\dag}_i,\ h^{\dag}_i=b_i$. The operators $h^{\dag}_i$ and $h_i$ 
are the creation and annihilation operator for holes. The previous 
symmetry has important consequences since it implies 
that the off-diagonal elements of the one-particle density matrix for 
$N_b$ HCB's [$\rho_{ij}(N_b)$] and for $N-N_b$ HCB's [$\rho_{ij}(N-N_b)$] are 
equal. Only the diagonal elements change, and they satisfy the 
relation $\rho_{ii}(N_b)=1-\rho_{ii}(N-N_b)$.

In the momentum distribution function ($n_k$), the particle-hole symmetry
leads to
\begin{equation}
\label{MomHCB}
n_k(N_b)= n_{-k}(N-N_b)+\left( 1-\frac{N-N_b}{N/2}\right),
\end{equation} 
so that an explicit dependence on the density appears. 
This behavior is different from the one of noninteracting 
spinless fermions, where the Hamiltonian is also particle-hole symmetric 
but under the transformation $h'_i= \left(-1\right)^i f^{\dagger}_i,\  
h'^{\dagger}_i=\left(-1\right)^i f_i$. 
In this case the momentum distribution function 
for fermions [$n^f_k(N_f)$] and [$n^f_k(N-N_f)$] are related 
by the expression 
\begin{equation}
\label{MomFermGS}
n^f_k(N_f)= 1-n^{f}_{-k+\pi}(N-N_f),
\end{equation} 
where there is no explicit dependence on the density.

\begin{figure}[h]
\begin{center}
\includegraphics[width=0.48\textwidth,height=0.64\textwidth]
{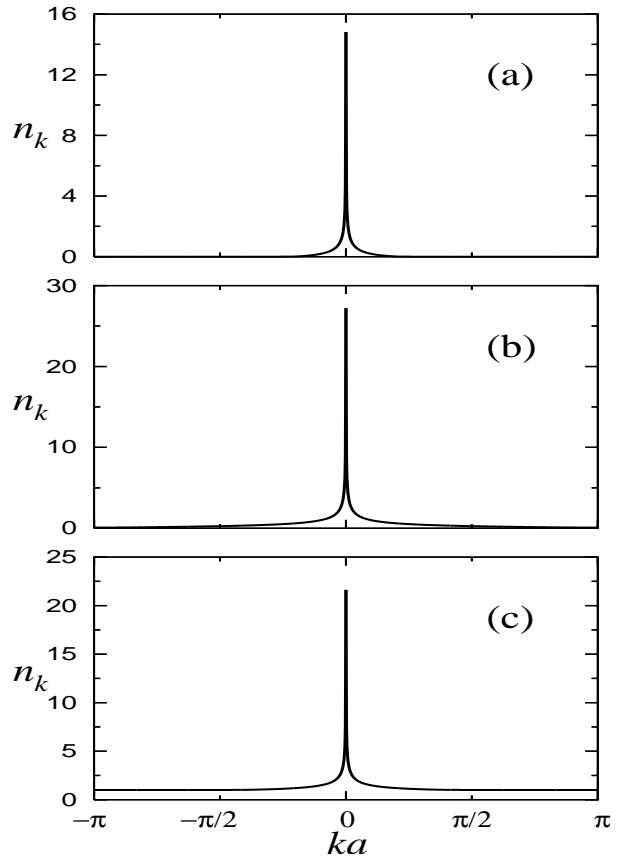}
\end{center} \vspace{-0.7cm}
\caption{Momentum profiles for periodic systems with 1000 lattice sites and 
occupations of 101 (a), 501 (b), and 799 (c) HCB's.}
\label{PerfilKHom}
\end{figure}
In Fig.\ \ref{PerfilKHom} we show the momentum profiles for systems with
three different fillings in 1000 lattice sites. Notice that we 
consider odd numbers of particles, which allows using periodic 
boundary conditions \cite{lieb61}. (In the case of an even number of 
particles antiperiodic boundary conditions are required \cite{lieb61}.) 
The peak structure in the momentum distribution function (Fig.\ 
\ref{PerfilKHom}) reflects the Bosonic nature of the particles, and 
is in contrast with the structure of the momentum distribution function 
for the equivalent noninteracting fermions. 
On increasing the number of HCB's two effects can be seen: 
(i) up to half filling the value of $n_{k=0}$ increases, and 
it starts to decrease when the number of particles exceeds half 
filling; (ii) the population of high momenta states is always increasing, 
showing that on increasing the density in the system, HCB's become more 
localized. The latter can be understood due to the impenetrability property 
of HCB's and the 1D character of the system.

\subsection{Off-diagonal correlations and $n_{k=0}$}

\begin{figure}[h]
\begin{center}
\includegraphics[width=0.47\textwidth,height=0.28\textwidth]
{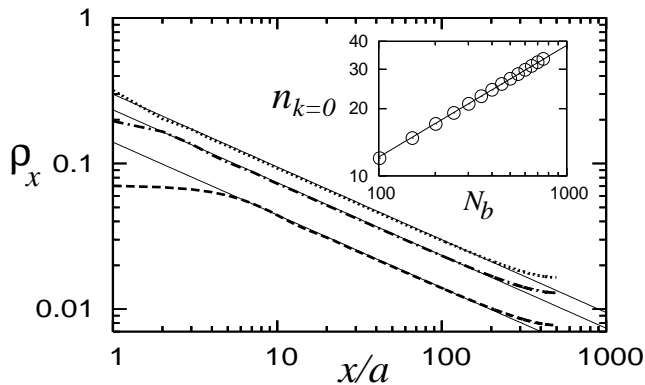}
\end{center} \vspace{-0.7cm}
\caption{One-particle density matrix $\rho_x$ vs $x/a$ for periodic systems 
with $N=1000$ and $\rho=0.5$ (dotted line), $\rho=0.211$ (dashed-dotted line), 
$\rho=0.071$ (dashed line). Thin continuous lines correspond to 
power laws $\sqrt{x/a}$. The inset shows $n_{k=0}$ vs $N_b$ 
for systems at half filling ($\bigcirc$), 
the line exhibits $\sqrt{N_b}$ behavior.}
\label{Largexhom}
\end{figure}

The formation of the peak in the momentum distribution function of the HCB's
is due to the off-diagonal quasi-long-range correlations present in the 
one-particle density matrix. As is shown in Fig.\ \ref{Largexhom} they 
decay as $\rho_x=A(\rho)/\sqrt{x/a}$ ($x=|x_i-x_j|$). $A(\rho)$ is a 
function of the density ($\rho=N_b/N$) in the system. A trivial calculation 
shows that in the thermodynamic limit, keeping the density constant, 
$n_{k=0}\sim 1/N\sum_{ij}|x_i-x_j|^{-1/2}$ scales as
\begin{equation}
n_{k=0}=B(\rho)\sqrt{N_b}=C(\rho)\sqrt{N}.
\label{nk0vsNbN}
\end{equation}
where $B(\rho)$ and $C(\rho)$ are functions of the density.
The previous result is valid to a good approximation for finite 
systems (for occupations higher than 100 HCB's), as shown 
in the inset in Fig.\ \ref{Largexhom} where we 
plot $n_{k=0}$ vs $N_b$ for different systems at half filling. 
The straight line displays a $\sqrt{N_b}$ behavior. 
 
\begin{figure}[h]
\begin{center}
\includegraphics[width=0.47\textwidth,height=0.29\textwidth]
{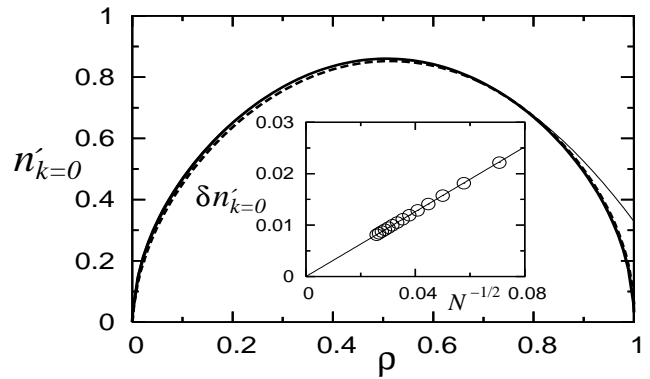}
\end{center} \vspace{-0.7cm}
\caption{Normalized occupation of $k=0$ ($n'_{k=0}$) as a function of the 
density in periodic systems with $N=1000$ (thick continuous line) and $N=300$ 
(dashed line). The thin continuous line shows a fit of 
Eq.\ (\ref{scalHom2}), with $b=1.51$ and $c=1.18$, to our numerical results 
for $N=1000$. The inset displays $\delta n'_{k=0}$ vs $N^{-1/2}$ (see text) 
for half filled systems ($\bigcirc$), the straight line shows 
the result of our fit to Eq.\ (\ref{finsizehom}).}
\label{PerfilKHomvsNb_S}
\end{figure}

A global picture of the occupation of the state with zero momentum 
as a function of the density 
is shown in Fig.\ \ref{PerfilKHomvsNb_S}. There we plot 
$n'_{k=0}=n_{k=0}/\sqrt{N}$ vs the density, and compare 
systems with 1000 and 300 lattice sites. The comparison shows that 
already for these system sizes the finite-size corrections are small. 
Actually, they are found to be given by 
\begin{equation}
n_{k=0}/\sqrt{N} =C(\rho)-D(\rho)/\sqrt{N}, \label{finsizehom}
\end{equation} 
where $D(\rho)$ can be positive or negative depending on the density. 
The inset in Fig.\ \ref{PerfilKHomvsNb_S} shows
$\delta n'_{k=0}=C(\rho)-n'_{k=0}$ vs $N^{-1/2}$ for 
half filled systems. The straight line displays the 
result of our fit for $C(\rho=0.5)$ and $D(\rho=0.5)$,
and confirms the validity of Eq.\ (\ref{finsizehom}).
The same functional form of the finite-size corrections 
was obtained for the continuous case \cite{forrester03},
which corresponds to systems in a lattice where 
the average interparticle distance is much larger than the lattice 
constant, i.e., in the limit $\rho=N_b/N\rightarrow 0$. 

\subsection{Low-density limit in the lattice}

In what follows we study the corrections introduced by the lattice to the 
occupation of the lowest momentum states in the continuous case. In a lattice 
at low momenta, the spectrum is quadratic in $k$ and $\delta k \sim 1/N$, 
so that the ratio between the level spacing and the band width reduces 
proportionally to $1/N^2$ when $N$ is increased. We find that the 
low-momentum occupations approach their values in the continuous case 
in exactly the same way,
\begin{equation}
\label{scalHom1}
\lambda_\eta\left(N_b,N\right) = \Lambda_{\eta} \left(N_b\right)
-\frac{E_\eta \left(N_b\right)}{N^2},
\end{equation} 
where $\lambda_\eta\left(N_b,N\right)\equiv n_{k=2\pi\eta/N a}\left(N_b\right)$ 
is the occupation of a momentum state $\eta$ when there are $N_b$ HCB's in a 
system with $N$ lattice sites. $\Lambda_{\eta} \left(N_b\right)$ is the  
occupation of the momentum state $\eta$ in the continuous case, and 
$E_\eta \left(N_b\right)$ for a given $\eta$ is a function of $N_b$. 

\begin{figure}[h]
\begin{center}
\includegraphics[width=0.47\textwidth,height=0.29\textwidth]
{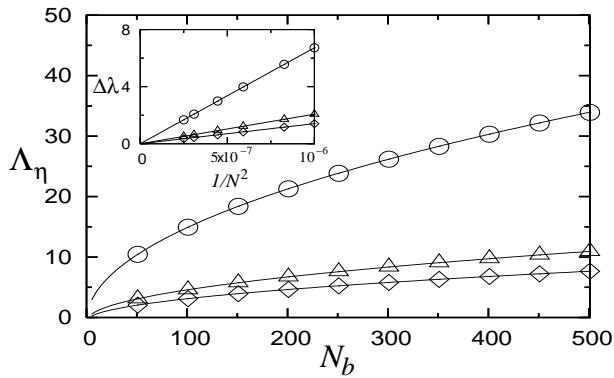}
\end{center} \vspace{-0.7cm}
\caption{Extrapolated values of the first ($\bigcirc$), 
second ($\triangle$), and third ($\Diamond$) 
momentum-state occupations in the continuous case ($\Lambda_\eta$). 
The lines following the data correspond to the results obtained in Ref.\ 
\cite{forrester03}. The inset shows $\Delta \lambda$ (see text) 
vs $1/N^2$ also for the first ($\bigcirc$), second ($\triangle$), 
and third ($\Diamond$) momenta states in systems with 501 HCB's and 
sizes between 1000 and 2000 lattice sites, the straight lines 
are the result of our fits.}
\label{PerfilKHomvsN1}
\end{figure}
We obtain $\Lambda_{\eta} \left(N_b\right)$ and $E_\eta \left(N_b\right)$ 
for the three lowest momenta states analyzing systems with fillings up to 
501 particles and sizes up to 2000 lattice sites. 
In Fig.\ \ref{PerfilKHomvsN1}, we compare our results for 
$\Lambda_{\eta} \left(N_b\right)$ with the ones in the continuous case 
presented in Ref.\ \cite{forrester03} [see Eqs.\ (56)--(58) there]. 
The agreement is excellent. In the inset we plot 
$\Delta \lambda=\Lambda_{\eta} \left(N_b\right)
-\lambda_\eta\left(N_b,N\right)$ also for the three lowest 
momentum states as a function of $1/N^2$ confirming Eq.\ (\ref{scalHom1}). 
Surprisingly, we find that Eq.\ (\ref{scalHom1}) works extremely well up 
to very high densities ($n\sim 0.8$). This feature allows one to understand 
the dependence of the lowest natural orbital occupation on the density 
(Fig.\ \ref{PerfilKHomvsNb_S}) as follows. 

The dependence of $\Lambda_{0} \left(N_b\right)$  
and $E_0 \left(N_b\right)$ on $N_b$ can be determined given 
Eq.\ (\ref{nk0vsNbN}) to be
\begin{equation}
\Lambda_{0} \left(N_b\right)= b\ \sqrt{N_b}, \qquad
E_0 \left(N_b\right)=c\ N_b^{5/2},
\end{equation} 
where $b$ and $c$ are two constants that can be obtained 
numerically. (Both relations above were confirmed by our numerical results.) 
Then Eq.\ (\ref{scalHom1}) may be rewritten as
\begin{equation}
\lambda_0\left(N,\rho\right)/\sqrt{N} = 
\sqrt{\rho} \left( b -c\ \rho^2 \right).\label{scalHom2}
\end{equation}
A fit of Eq.\ (\ref{scalHom2}) to our numerical results for $N=1000$, 
with $b=1.51$ and $c=1.18$, is shown in 
Fig.\ \ref{PerfilKHomvsNb_S} as a thin continuous line. 
As it can be seen, Eq.\ (\ref{scalHom2}) describes very well the numerical 
results up to densities $n\sim 0.8$, which were the densities up to which 
we had found that Eq.\ (\ref{scalHom1}) was a good approximation.

\subsection{$n_{k=0}$ singularity in the thermodynamic limit}

\begin{figure}[h]
\begin{center}
\includegraphics[width=0.47\textwidth,height=0.27\textwidth]
{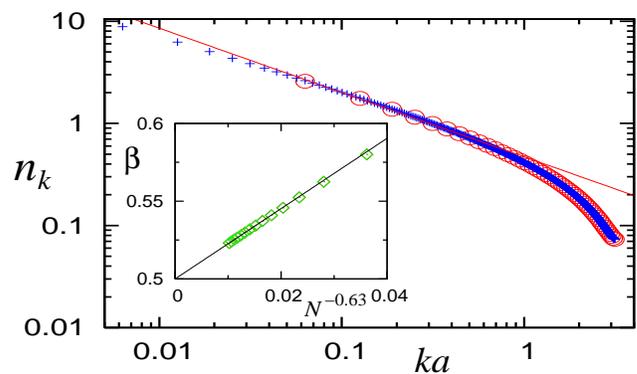}
\end{center} \vspace{-0.7cm}
\caption{(Color online) Log-log plot of the momentum distribution function of 
systems at half filling with $N=100$ (\textcolor{red}{$\bigcirc$}) and $N=1000$ 
(\textcolor{blue}{$+$}) lattice sites. The straight line shows a power-law fit 
$n_k\sim k^{-\beta}$ ($\beta=0.63$) to the lowest momentum states of the 
system with $N=100$. The inset displays the dependence of the exponent 
$\beta$ (fitted for half filled systems) on the system size. 
The straight line shows that it extrapolates to $\beta=0.5$ for 
$N\rightarrow\infty$.}
\label{PerfilKHomHalfFill}
\end{figure}

To conclude this section on periodic systems, we analyze in detail the 
low-momentum region of the momentum distribution function. The power-law 
decay of the one-particle density matrix ($\rho_x\sim 1/\sqrt{x/a}$) implies 
that in the thermodynamic limit the momentum distribution function has a 
$|k|^{-\beta}$ ($\beta=1/2$) singularity at $k=0$ \cite{vaidya79}. For 
finite systems we find that, due to finite-size effects, the apparent exponent 
observed in the low-momentum region of $n_k$ depends strongly on the number 
of particles in the system. In Fig.\ \ref{PerfilKHomHalfFill} we show a 
log-log plot of the momentum distribution of two systems at half filling 
with $N=100$ and $N=1000$ lattice sites. A fit to a power law in the case of 
100 lattice sites reveals an exponent $\beta=0.63$. As the figure shows, the 
increase of the system size reduces the exponent of the power law only in the 
low-momentum region. We have studied systems at half filling fitting power 
laws from the lowest ten momentum states available (excluding $k=0$). The 
results are shown in the inset of Fig.\ \ref{PerfilKHomHalfFill}. This inset 
shows that with increasing the system size the exponent $\beta$ extrapolates 
very slowly to 0.5. Basically, we find that it extrapolates as a power law 
$\sim N^{-0.63}$. The latter exponent is a nonuniversal one since it depends 
on the way in which one fits power laws for the low-momentum region of $n_k$. 
Up to the bigger system we calculated $N=1500$, the exponent $\beta$ reduced 
only up to 0.523. Our results imply that the experimental observation of 
the thermodynamic limit exponent $\beta=0.5$ will be very difficult if not 
impossible considering finite-temperature effects \cite{paredes04}.

\section{Hard-core bosons confined in harmonic traps \label{HCBGSIII}}

We analyze in this section the case in which HCB's are confined in 
harmonic traps. The Hamiltonian in this case is given by Eqs.\ 
(\ref{HamHCB}) and (\ref{ConstHCB}) with $\alpha=2$, 
and is not particle-hole symmetric like in the periodic case.
In order to quantitatively characterize these systems we make 
use of the length scale set by the combination lattice-confining 
potential $\zeta=\left(V_{\alpha}/t \right)^{-1/2}$, 
and the associated characteristic density $\tilde{\rho}=N_b/\zeta$, 
both introduced in Refs.\ \cite{rigol03_1,rigol03_2,rigol03_3}. 
Since the diagonal elements of the one-particle density matrix for HCB's 
and the equivalent fermions are equal [see Eq.\ (\ref{green1}) for $i=j$], 
density properties of both systems trivially coincide. We concentrate here 
in the quantities related to the off-diagonal correlations of the HCB's, 
like the momentum distribution function and the natural orbitals.

The natural orbitals ($\phi^\eta_i$) are defined as the eigenfunctions
of the one-particle density matrix \cite{penrose56}, 
\begin{equation}
\label{NatOrb}
\sum^N_{j=1} \rho_{ij}\phi^\eta_j=
\lambda_{\eta}\phi^\eta_i,
\end{equation}
and can be understood as being effective single-particle states with 
occupations $\lambda_{\eta}$. The lowest natural orbital (the highest 
occupied one) is considered to be the condensate. In the periodic case 
the natural orbitals are plane waves, i.e, the eigenfunctions of 
Eq.\ (\ref{NatOrb}) are momentum states (the lowest natural orbital 
is the state with momentum zero). On the contrary, since in the trapped 
case the translational invariance is broken, the natural orbitals 
occupations and the momentum distribution function do not coincide.

\begin{figure}[h]
\begin{center}
\includegraphics[width=0.49\textwidth,height=0.52\textwidth]
{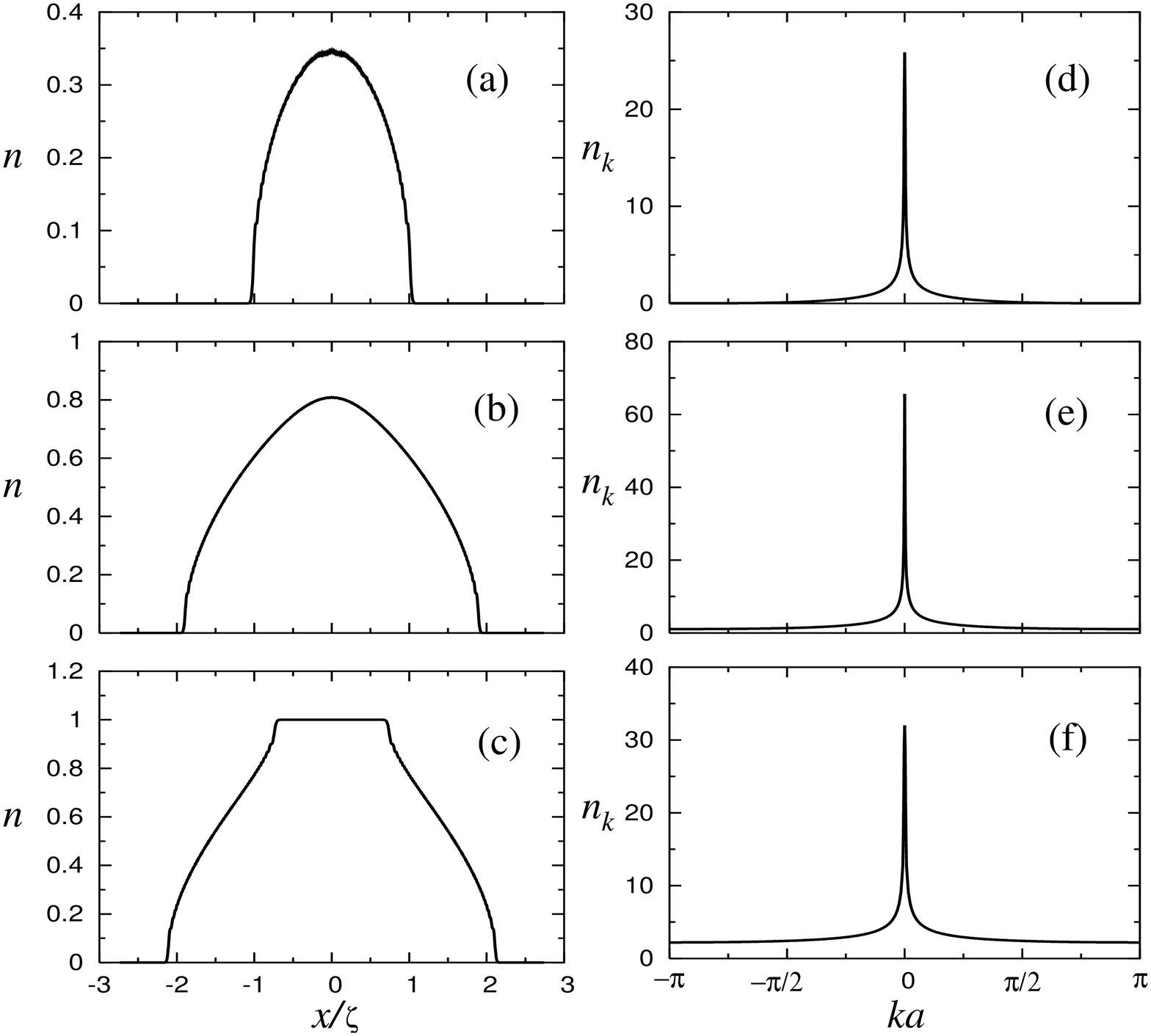}
\end{center} \vspace{-0.7cm}
\caption{Density (a)--(c) and normalized momentum distribution function 
(d)--(f) for trapped systems with 1000 lattice sites, 
$V_2a^2=3\times 10^{-5}t$ and occupations of 101 (a),(d), 401 (b),(e), 
and 551 (c),(f) HCB's.}
\label{Perfiles1000HCB}
\end{figure}
In Fig.\ \ref{Perfiles1000HCB} we show three density profiles and their 
corresponding normalized momentum distribution function for traps with 1000 
lattice sites. Positions are normalized by the characteristic length $\zeta$, 
and the normalized momentum distribution function is defined as 
$n_k=(a/\zeta)\sum^N_{ij=1} e^{-ik(i-j)}\langle b^\dagger_{i}b_{j}\rangle$. 
Figure \ref{Perfiles1000HCB} displays features similar to the periodic 
case. The normalized momentum distribution function exhibits narrow peaks 
at $k=0$ with $n_{k=0}$ initially increasing with the number of particles, 
and decreasing after certain filling when more particles are added to 
the system. Localization effects also start to be evident at high 
fillings, with an increment of the population of high momenta states 
[Fig.\ \ref{Perfiles1000HCB}(f)], and the formation of Fock states 
(or a Mott insulating plateau) in the middle of the system 
[sites with $n_i=1$ in Fig.\ \ref{Perfiles1000HCB}(c)]. Notice that 
in the latter case the formation of the Mott insulator did not destroy
the sharp peak observed at $n_{k=0}$, which is associated to the 
superfluid phase surrounding the Mott plateau.

\begin{figure}[h]
\begin{center}
\includegraphics[width=0.48\textwidth,height=0.6\textwidth]
{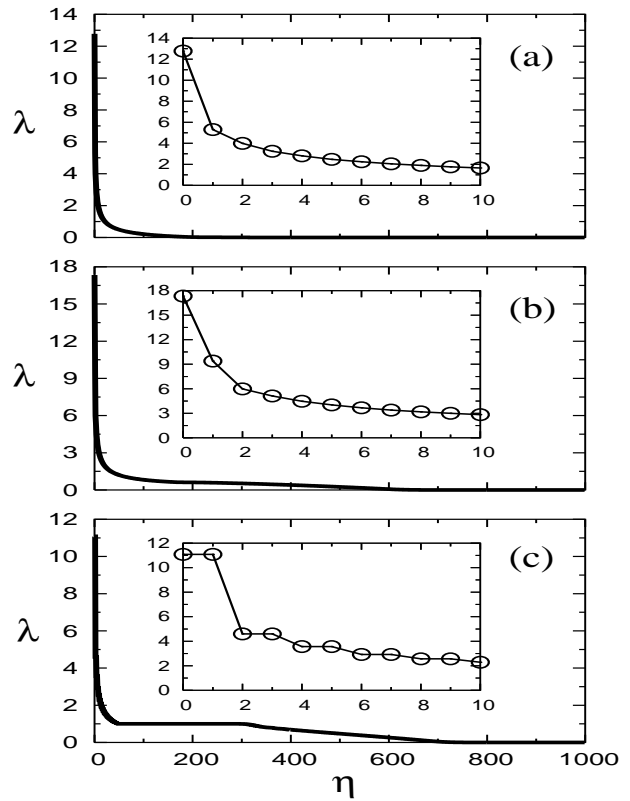}
\end{center} \vspace{-0.7cm}
\caption{Occupation of the natural orbitals for 
trapped systems with 1000 lattice sites, $V_2a^2=3\times 10^{-5}t$, 
and occupations of 101 (a), 401 (b), and 551 (c) HCB's. 
The insets show the occupation of the lowest eleven natural 
orbitals in each case.}
\label{NatOrb1000}
\end{figure}
Results obtained for the occupations of the natural orbitals, in the systems 
of Fig.\ \ref{Perfiles1000HCB}, are presented in Fig.\ \ref{NatOrb1000}.
The occupations are plotted as a function of the orbital numbers $\eta$, 
and they are ordered starting from the highest one. The effects of 
increasing the filling in the system are similar to the ones observed
in the momentum distribution function. The occupation of the lowest natural 
orbital increases with increasing the number of particles up to a certain 
filling where it starts to decrease when more particles are added. In 
contrast to the momentum distribution function, the natural orbital 
occupations exhibit clear signatures of the formation of the Mott insulating 
state in the middle of the system since a plateau with $\lambda=1$ appears 
[Fig.\ \ref{NatOrb1000}(c)], and the natural orbitals with $\lambda\neq 1$ 
become pairwise degenerated [inset in Fig.\ \ref{NatOrb1000}(c)]. This can be 
easily understood since in the region where $n_i=1$, the one-particle density 
matrix is diagonal $\rho_{ij}=\delta_{ij}$, so that it can be seen as a 
three block diagonal matrix with the block in the middle being an 
identity matrix and the other two being identical. Thus the 
diagonalization gives a group of eigenvalues $\lambda_\eta=1$ and all the 
others pairwise degenerated.

\begin{figure}[h]
\begin{center}
\includegraphics[width=0.49\textwidth,height=0.57\textwidth]
{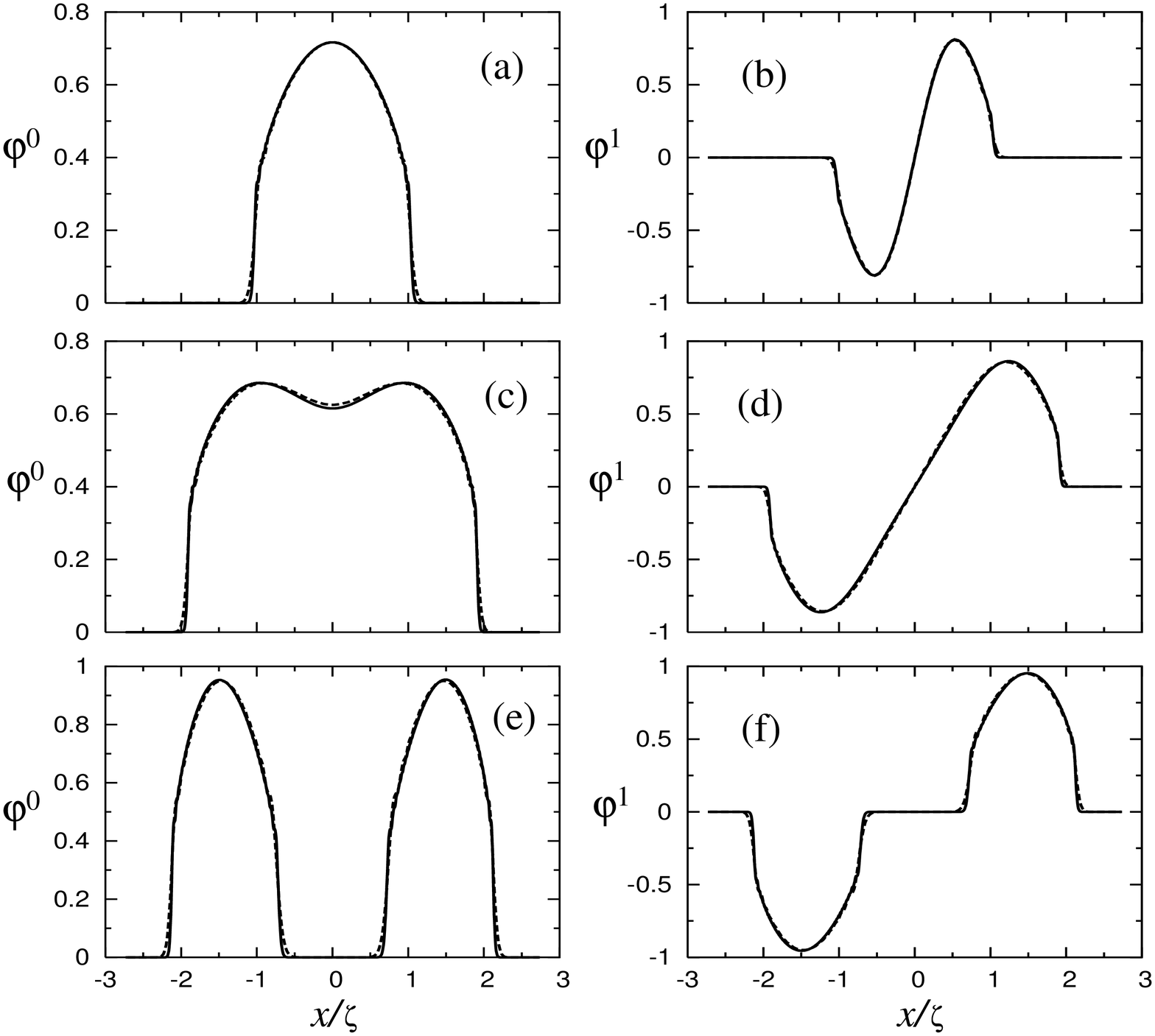}
\end{center} \vspace{-0.7cm}
\caption{Profiles of the two lowest natural orbitals for 
trapped systems with: 1000 lattice sites, 
$V_2a^2=3\times 10^{-5}t$, and occupations of 101 (a),(b), 401 (c),(d), and 
551 (e),(f) HCB's (continuous line); 300 lattice sites, $V_2a^2=3.3\times 
10^{-4}t$, and occupations of 30 (a),(b), 121 (c),(d), and 
167 (e),(f) HCB's (dashed line).}
\label{NatOrbWF1000}
\end{figure}
Profiles of the two lowest natural orbitals, for the same 
parameters of Figs.\ \ref{Perfiles1000HCB} and \ref{NatOrb1000}, 
are presented in Fig.\ \ref{NatOrbWF1000}. Positions in the trap
are also normalized by $\zeta$ and the proper definition of the
scaled natural orbitals is given by 
\begin{equation}
\varphi^\eta=R^{1/2} \phi^\eta, \qquad 
R=\left( N_b \zeta/ a\right)^{1/2}. \label{NOSGS}
\end{equation}
This scaling relation is valid only for the lowest natural orbitals and is 
meaningful for comparing systems with the same characteristic density. 
Notice that for a given $\tilde{\rho}$ the occupied system size ($L$) 
is proportional to $\zeta$, i.e, $L=F(\tilde{\rho}) \zeta$ 
with $F(\tilde{\rho})$ depending only on the value of $\tilde{\rho}$.
In Fig.\ \ref{NatOrbWF1000} we have also plotted
results for smaller systems (dashed line) but fulfilling 
the condition of equal characteristic density. The comparison between 
the plots shows that the scaling relation defined by 
Eq.\ (\ref{NOSGS}) holds. (We find this to be valid to a good 
approximation for occupied system sizes larger than 100 lattice sites.) 
In Figs.\ \ref{NatOrbWF1000}(a) and (b) it can be seen that for low 
characteristic densities the natural orbitals are similar to 
the ones of systems without the optical lattice 
\cite{papenbrock03,forrester03}. 
On increasing the characteristic density the shape of the lowest 
natural orbital starts to change since its weight starts to reduce in the 
middle of the trap, as shown in Fig.\ \ref{NatOrbWF1000}(c). Once the 
Mott insulating plateau appears in the center of the system, degeneracy sets 
in for the lowest natural orbital [inset in Fig.\ \ref{NatOrb1000}(c)], 
and their weight in the region with $n=1$ vanishes [Figs.\ 
\ref{NatOrbWF1000}(e) and (f)].

\subsection{Off-diagonal correlations and the lowest natural orbital 
occupation}

\begin{figure}[h]
\begin{center}
\includegraphics[width=0.47\textwidth,height=0.28\textwidth]
{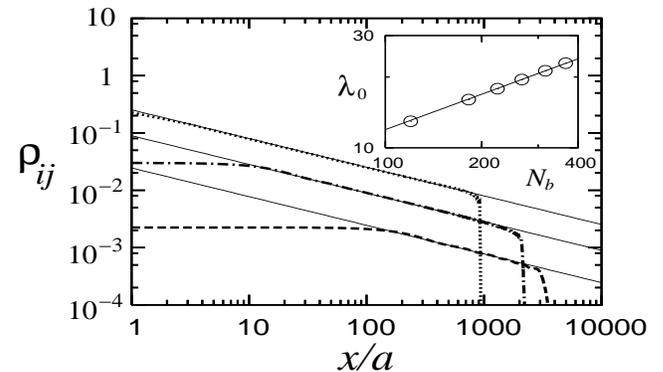}
\end{center} \vspace{-0.7cm}
\caption{HCB one-particle density matrix vs $x/a$ ($x = \mid x_i -x_j \mid$) 
for systems with $N_b=1000$, $\tilde{\rho}=2.0$, $n_i=0.75$ (dotted line),  
$N_b=100$, $\tilde{\rho}=4.47\times 10^{-3}$, 
$n_i=0.03$ (dashed-dotted line), and $N_b=11$, 
$\tilde{\rho}=2.46\times 10^{-5}$, $n_i=2.3\times 10^{-3}$ (dashed line). 
The point $i$ was chosen in the middle of the trap and only $j$ was changed. 
The abrupt reduction of $\rho_{ij}$ occurs for $n_j\rightarrow 0$. Thin 
continuous lines correspond to power laws $\sqrt{x/a}$. The inset shows 
$\lambda_0$ vs $N_b$ for systems with $\tilde{\rho}=1.0$ ($\bigcirc$). 
The straight line exhibits $\sqrt{N_b}$ behavior.}
\label{Largextrap}
\end{figure}

The peaks appearing in the momentum distribution function at $k=0$, 
and the high occupation of the lowest natural orbitals are consequences 
of the off-diagonal quasi-long-range correlations present in the 
one-particle density matrix ($\rho_{ij}$). We obtain that away from 
$n_i,n_j=0,1$, $\rho_{ij}$ decays as a power law 
$\rho_{ij}\sim |(x_i-x_j)/a|^{-1/2}$ for large distances 
independently of the fact that the density is changing in the system. 
This is shown in Fig.\ \ref{Largextrap} where we plot $\rho_{ij}$ for
traps with fillings ranging over three decades. As an approximation, 
one may consider that $\rho_{ij}$ and $|(x_i-x_j)/a|^{-1/2}$ are related 
through a constant that depends only on the characteristic density 
$\tilde{\rho}$, which in trapped systems plays a role similar to that 
played by the density in periodic systems. 

\begin{figure}[h]
\begin{center}
\includegraphics[width=0.47\textwidth,height=0.28\textwidth]
{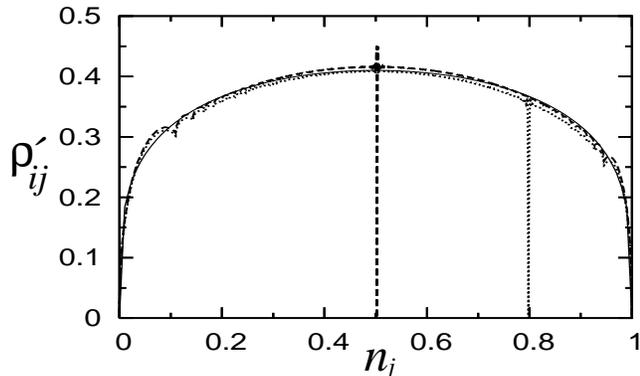}
\end{center} \vspace{-0.7cm}
\caption{HCB one-particle density matrix $\rho'_{ij}$ 
[as in Eq.\ (\ref{rhop})] vs $n_j$. $\rho'_{ij}$ is measured 
from points $i$ with $n_{i}=0.5$ (dashed line) and 
$n_{i}=0.8$ (dotted line) in a trap with $N_b=951$, 
$\tilde{\rho}=2.61$ so that a Mott insulator is formed in the middle 
of the system. The thin continuous line corresponds to 
$0.58[n_j(1-n_j)]^{1/4}$.}
\label{Largexvsrho}
\end{figure}

A detailed study shows, however, that $\rho_{ij}$ and 
$|(x_i-x_j)/a|^{-1/2}$ are related by a slowly varying 
(away from $n_i,n_j= 0,1$) function of the density, 
which we denote as $f_2(n_i,n_j)$. We find this function to be 
\begin{equation}
\label{f2}
f_2(n_i,n_j) \sim [n_i(1-n_i)n_j(1-n_j)]^{1/4}.
\end{equation}  
In order to show it, we plot in Fig.\ \ref{Largexvsrho}
\begin{equation}
\label{rhop}
\rho'_{ij}=\rho_{ij}|(x_i-x_j)/a|^{1/2}/[n_i(1-n_i)]^{1/4}
\end{equation} 
vs $n_j$ for two different values of $n_i$ in a system where a Mott 
insulator is formed in the middle of the trap. (In this case the density 
in the superfluid phase changes between 0 and 1.) 
The results for $\rho'_{ij}$ are then 
compared with a function $A[n_j(1-n_j)]^{1/4}$, where as a fit we obtain 
$A=0.58$. The agreement between the plots shows that $f_2(n_i,n_j)$ in 
Eq.\ (\ref{f2}) describes very well the density dependence of the 
one-particle density matrix in harmonically trapped systems.

\begin{figure}[h]
\begin{center}
\includegraphics[width=0.47\textwidth,height=0.28\textwidth]
{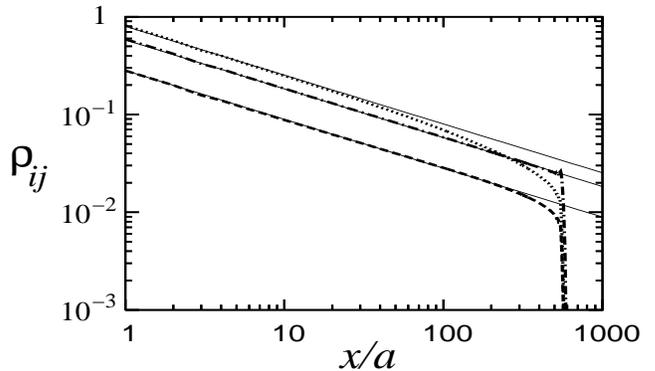}
\end{center} \vspace{-0.7cm}
\caption{One-particle density matrix normalized in different ways 
(for the same system) vs $x/a$ ($x = \mid x_i -x_j \mid$): 
$\rho_{ij}$ (dashed line), $\rho^N_{ij}$ as in Eq.\ (\ref{rhoN}) 
(dashed-dotted line), and $\rho^K_{ij}$ as in Eq.\ (\ref{rhoK}) (dotted line). 
Thin continuous lines exhibit $|x_i -x_j|^{-1/2}$ behavior. The plots were made 
for trap with $N_b=951$ and $\tilde{\rho}=2.61$ so that a Mott insulator is 
formed in the middle of the system. In addition, $i$ was fixed ($n_{i}=0.34$), 
and $j$ was changed with $n_j>n_i$.}
\label{LargexrhoInset}
\end{figure}

The fact that the dependence of $\rho_{ij}$ on the density is very 
small (a fourth root of the densities) is what allows one to observe 
the $|(x_i-x_j)/a|^{-1/2}$ behavior without the need of normalizing 
the one-particle density matrix. For very low characteristic 
densities ($n_i,n_j\rightarrow 0$) the function $f_2(n_i,n_j)$ 
reduces to the exact result known for the continuous case 
$f_2(n_i,n_j)\sim (n_in_j)^{1/4}$ \cite{forrester03,gangardt04}. 
In Fig.\ \ref{LargexrhoInset} we compare results obtained 
for $\rho_{ij}$ and its normalized version,
\begin{equation}
\rho^N_{ij}=\rho_{ij}/[n_i(1-n_i)n_j(1-n_j)]^{1/4}.
\label{rhoN}
\end{equation} 
The difference between the power-law decay in both cases is small and can 
be only distinguished at very large distances. In addition, we have also 
plotted the results obtained for $\rho_{ij}$ normalized following Kollath 
{\it et~al.} \cite{kollath04},
\begin{equation}
\label{rhoK}
\rho^K_{ij}=\rho_{ij}/\sqrt{n_i n_j}.
\end{equation} 
As Fig.\ \ref{LargexrhoInset} shows for the HCB case, 
a possible power-law behavior in $\rho^K_{ij}$ disappears much 
before the one observed in $\rho_{ij}$ without normalization. 
This implies that the scaling of the correlations chosen in 
Ref.\ \cite{kollath04} does not apply to very strong Hubbard repulsions.

Considering the results for the power-law decay of the one-particle 
density matrix and for the scaling of the lowest natural orbital, it is 
possible to calculate how the lowest natural orbitals occupation scales 
in the thermodynamic limit ($\lambda_0=\sum_{ij}\phi^0_i \rho_{ij}\phi^0_j$) 
defined by a constant characteristic density. 
Replacing the sums in $\lambda_0$ by integrals ($L \gg a$), 
{\setlength\arraycolsep{0.0pt}
\begin{eqnarray}
&&\lambda_0 \sim 1/a^2
\int^L_{-L}dx \int^L_{-L}dy 
\frac{\phi^0(x)f_2[n(x),n(y)]\phi^0(y)}{|(x-y)/a|^{1/2}} 
\nonumber \\ &&= \left( \zeta/a\right) ^{3/2}R^{-1}
\begin{array}{l}\ \ \ F(\tilde{\rho})\\ 
\ \ \ \int\int\\ -F(\tilde{\rho})\end{array}
dX dY \frac{\varphi^0(X)f_2[n(X),n(Y)]\varphi^0(Y)}{|X-Y|^{1/2}}
\nonumber \\ &&= B_2(\tilde{\rho})\sqrt{N_b}=
C_2(\tilde{\rho}) \sqrt{\zeta/a} \ , \label{lambda0HCBGS}
\end{eqnarray}}\noindent where we did the change of variables 
$x=X \zeta$, $y=Y \zeta$, and 
$\phi^0=R^{-1/2}\varphi^0$. The integral over $X,Y$ 
depends only on the characteristic density. 
This is so because we have shown before 
that the scaled natural orbitals as a function of $x/\zeta$ do not 
change when the characteristic density is kept constant. In addition, 
also the density profiles (as a function of $x/\zeta$) do not change 
when the characteristic density is kept constant. This feature, which is 
valid for HCB's and spinless fermions, was proven in Ref.\ \cite{rigol03_2}.
Hence $B_2(\tilde{\rho})$ and $C_2(\tilde{\rho})$ 
depend only on $\tilde{\rho}$. The previous results show that with the 
properly defined characteristic density, the occupation 
of the lowest natural orbital scales like in the homogeneous case, 
proportionally to $\sqrt{N_b}$. Equation (\ref{lambda0HCBGS}) is valid 
to a good approximation for finite systems, as can be seen in the 
inset in Fig.\ \ref{Largextrap} where we plot $\lambda_0$ 
vs $N_b$ for different systems with characteristic density 
$\tilde{\rho}=1$, the straight line displays a $\sqrt{N_b}$ behavior.

\begin{figure}[h]
\begin{center}
\includegraphics[width=0.49\textwidth,height=0.31\textwidth]
{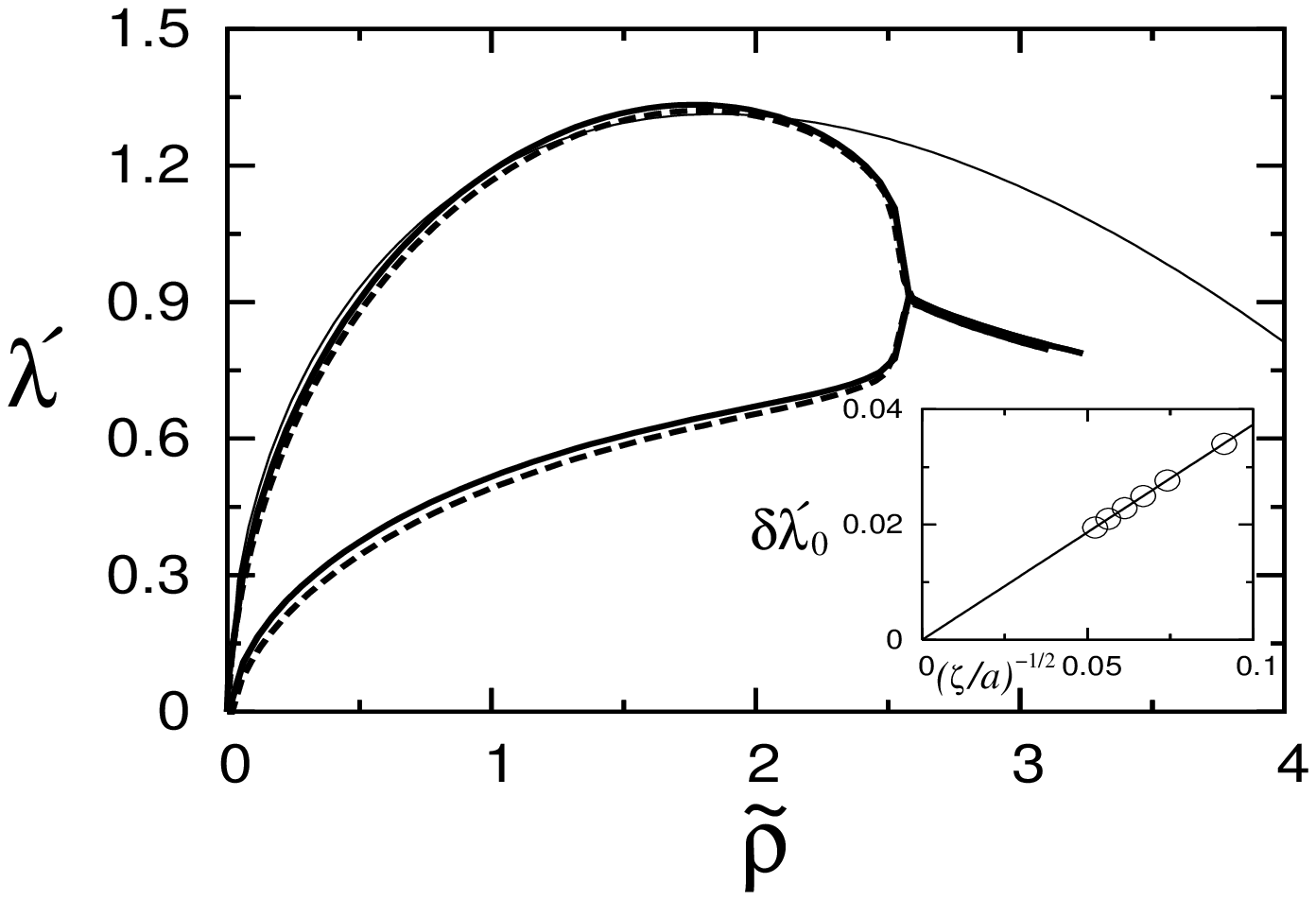}
\end{center} \vspace{-0.7cm}
\caption{Normalized occupation of the two lowest natural orbitals (see text) 
vs $\tilde{\rho}$. The systems analyzed have $N=1000$, 
$V_2a^2=3\times 10^{-5}t$, and occupations up to 600 particles 
(thick continuous line), and $N=300$, $V_2a^2=3.3\times 10^{-4}t$, 
and occupations up to 270 particles (dashed line). 
The thin continuous line shows a fit of Eq.\ (\ref{scalTrap2}), 
with $b_2=1.44$ and $c_2=0.26$, to our numerical results for $N=1000$. 
The inset displays $\delta \lambda'_0$ vs $N^{-1/2}$ (see text) for 
traps with $\tilde{\rho}=1$ ($\bigcirc$), the straight line 
shows the result of our fit to Eq.\ (\ref{finsizetrap}).}
\label{NatOrbTrapvsNb}
\end{figure}
A global picture of the occupation of the two lowest natural orbitals 
is shown in Fig.\ \ref{NatOrbTrapvsNb}. There we plot 
$\lambda'=\lambda/\sqrt{\zeta/a}$ vs $\tilde{\rho}$, 
and compare systems with different curvatures of the confining potential. 
The Mott insulator forms in the middle of the trap when the degeneracy 
appears in the natural orbitals. The comparison between the plots shows that 
already for these system sizes (larger than 100 lattice sites) 
the finite-size corrections are small. 
They were found to be determined by the expression 
\begin{equation}
\lambda_0/\sqrt{\zeta/a}= 
C_2(\tilde{\rho})-D_2(\tilde{\rho})/\sqrt{\zeta/a},
\label{finsizetrap}
\end{equation} 
with $D_2(\tilde{\rho})$ being a function of the characteristic density. 
Equation (\ref{finsizetrap}) has the same form as Eq.\ (\ref{finsizehom}) 
for the homogeneous case, when $N$ is substituted by $\zeta/a$. 
In the inset in Fig.\ \ref{NatOrbTrapvsNb}, we plot
$\delta \lambda'_0=C_2(\tilde{\rho})-\lambda'_0$ vs $N^{-1/2}$ for 
systems with $\tilde{\rho}=1$. The straight line displays the 
result of our fit for $C_2(\tilde{\rho}=1)$ and $D_2(\tilde{\rho}=1)$,
and confirms the validity of Eq.\ (\ref{finsizetrap}).
In the case without the lattice \cite{forrester03}, the same functional 
form was obtained for the finite-size corrections in terms of the number 
of HCB's in harmonic traps. 

\subsection{Formation of the Mott insulator}

\begin{figure}[h]
\begin{center}
\includegraphics[width=0.48\textwidth,height=0.30\textwidth]
{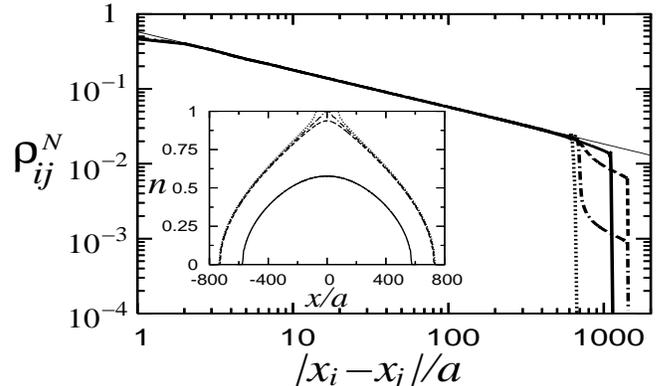}
\end{center} \vspace{-0.7cm}
\caption{HCB one-particle density matrix $\rho^N_{ij}$ [Eq.\ (\ref{rhoN})] 
on approaching the Mott insulating state. In the plots the 
trap curvature was kept constant $V_2a^2=7.5\times10^{-06}t$, and the 
filling was increased: $N_b=500$ (thick continuous line), 
$N_b=911$ (dashed line), $N_b=931$ (dashed-doted line), 
and $N_b=951$ (dotted line). The corresponding density profiles are shown 
in the inset. The thin continuous line following $\rho^N_{ij}$ corresponds 
to a power law $0.58(|x_i-x_j|/a)^{-1/2}$. $\rho^N_{ij}$ was measured in the 
four cases displayed fixing $i$ so that $n_i\sim 0.2$.}
\label{LargexMI}
\end{figure}

So far we have analyzed generic features of harmonically trapped systems 
in a lattice. In what follows we study in more detail the formation of 
the Mott-insulating state in the middle of the trap. In Fig.\ 
\ref{Largextrap} we have shown the behavior of the one-particle density 
matrix in systems with fillings below the one at which the Mott insulator 
appears in the middle of the trap, and once the Mott insulator has formed, 
splitting the trap in two parts. In all these cases power-law decays 
$\rho_{ij}\sim |x_i-x_j|^{-1/2}$ are observed. We find that when the local 
density approaches to one in the middle of the trap, the power law 
$\rho_{ij}\sim |x_i-x_j|^{-1/2}$ observed in one half of the system starts 
to disappear on entering in the other half. This effect can be seen in 
Fig.\ \ref{LargexMI} where we plot $\rho^N_{ij}$ [Eq.\ (\ref{rhoN})] vs 
$|x_i-x_j|$ (for fixed $i$ in one side of the trap) when the density in 
the middle of the system $n\rightarrow 1$. The density profiles 
corresponding to these density matrices are presented 
in the inset. The fast decay of the one-particle density matrix when crossing 
the middle of the trap (for $n\rightarrow 1$ in the center of the system) 
leads to the formation of two independent and identical 
quasicondensates in each side of the trap. This is reflected in 
Fig.\ \ref{NatOrbTrapvsNb} since the occupation of the two lowest natural 
orbitals starts to be similar on approaching the formation of the Mott 
insulating state.

\begin{figure}[h]
\begin{center}
\includegraphics[width=0.47\textwidth,height=0.28\textwidth]
{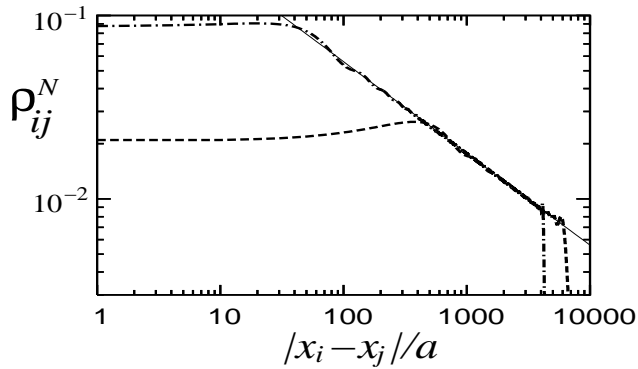}
\end{center} \vspace{-0.7cm}
\caption{HCB one-particle density matrix $\rho^N_{ij}$ [Eq.\ (\ref{rhoN})] 
for very dilute systems with $N_b=100$, $\tilde{\rho}=4.47\times 10^{-3}$ 
(dashed-dotted line), and $N_b=11$, $\tilde{\rho}=2.46\times 10^{-5}$ 
(dashed line). The thin continuous line following $\rho^N_{ij}$ corresponds 
to a power law $0.56(|x_i-x_j|/a)^{-1/2}$. $\rho^N_{ij}$ was measured  
in both cases displayed fixing $i$ in one border of the trap.}
\label{Largexcont}
\end{figure}

In Fig.\ \ref{LargexMI} we have also plotted a case where the density in 
the middle of the system is $n\sim 0.58$ (thick continuous line). 
In this case one can observe very small deviations from the $|x_i-x_j|^{-1/2}$ 
behavior in one half of the system on entering in the other half. This shows
that although the destruction of quasi-long-range correlations between the 
two halves of the trap occurs only when the density in the middle of the system
approaches $n_i=1$, deviations from the power law $|x_i-x_j|^{-1/2}$ 
are observed before. They just increase with increasing the density in the 
middle of the trap. Only in the case of very low characteristic 
densities is the power law $|x_i-x_j|^{-1/2}$ observed in the whole 
system as shown in Fig.\ \ref{Largexcont}. 
A comparison between Figs.\ \ref{LargexMI} and \ref{Largexcont} reveals 
an interesting feature, which is also present in the homogeneous case 
(Fig.\ \ref{Largexhom}). The interface between the short-distance and 
large-distance behavior of the one-particle density matrix moves toward 
shorter distances with increasing the density in the system.

\begin{figure}[h]
\begin{center}
\includegraphics[width=0.46\textwidth,height=0.29\textwidth]
{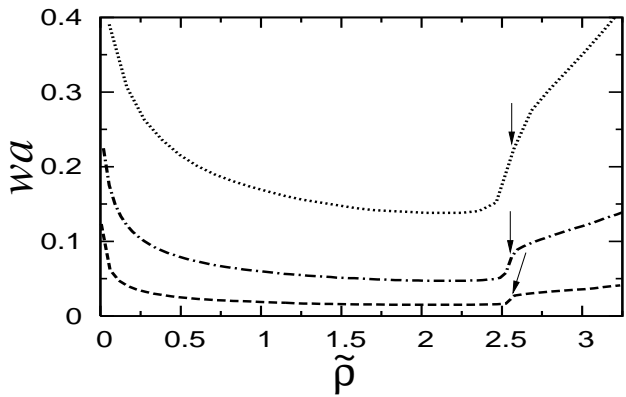}
\end{center} \vspace{-0.7cm}
\caption{Full width at half maximum of the momentum distribution 
function ($w$) as a function of the characteristic density for 
three different traps with $V_2a^2=3.0\times 10^{-3}t$ (dotted line), 
$V_2a^2=3.3\times 10^{-4}t$ (dashed-dotted line), and 
$V_2a^2=3.0\times 10^{-5}t$ (dashed line). 
The arrows signal the point at which the Mott insulator sets in the middle 
of the trap. In the plot $w$ was normalized by the lattice constant $a$.}
\label{HalfWidth}
\end{figure}

The destruction of the power-law decay of $\rho_{ij}$ observed in 
Fig.\ \ref{LargexMI} is reflected by an increase of the full width at 
half maximum of the momentum distribution function ($w$) close before 
the Mott insulator appears in the middle of the system. $w$ is 
proportional to the inverse of the correlation length in the system. 
In the HCB case the correlation length is zero in the Mott-insulating 
phase and proportional to the size of the region $0<n<1$ for the 
superfluid phase. This implies that on filling the trap $w$ starts 
to reduce (the system size increases) up to the point 
at which the system begins to split close before the formation of the Mott 
insulating state in the middle of the trap. This can be seen in 
Fig.\ \ref{HalfWidth} where we plot $w$ as a function of the characteristic 
density for different system sizes. The points at which the Mott insulator 
appears are signaled by an arrow ($\tilde{\rho}\sim 2.6 -- 2.7$). The figure 
also shows that, as it can be intuitively expected, the increase of 
$w$ becomes smaller on increasing the system 
size since the superfluid phase (the one producing the peak) 
becomes larger. 
The proposal of measuring $w$ in order to (approximately) 
detect the parameter region where the local Mott insulator appears in the 
trap was done by Kollath {\it et~al.} \cite{kollath04} for 1D systems 
and discussed by Wessel {\it et~al.} \cite{wessel04} for higher dimensional 
systems. As it was shown by recent experimental results \cite{stoferle03} 
for the systems sizes available experimentally $w$ gives a signal 
of when the Mott insulator appears for the soft-core boson case. 
Since local Mott-insulating and superfluid phases coexist, in order to 
unambiguously characterize these systems local probes are needed. This has 
been discussed in Refs.\ \cite{batrouni02,wessel04} for soft-core bosons 
and Refs.\ \cite{rigol03_1,rigol03_2} for fermions. The basic idea of 
local probes follows from the fact that the superfluid phases are 
compressible while the Mott insulating phases are incompressible 
(a gap opens in the charge excitations). However, such local probes have 
not been yet implemented experimentally. 

\subsection{Low-density limit in the lattice}

Finally, we study the corrections introduced by the lattice to the 
occupation of the lowest natural orbitals in the confined continuous 
case. Due to the combination lattice-harmonic potential the ratio 
between the level spacing and the bandwidth in the low-energy region 
is proportional to $(\zeta/a)^{-1}$. As in the periodic case, we find that
the occupation of the lowest natural orbitals approach their value in the 
continuous system $\Lambda_{\eta}\left(N_b\right)$ in the same way,
\begin{equation}
\label{scalTrap1}
\lambda_\eta\left(N_b,\zeta/a\right) = \Lambda_{\eta}\left(N_b\right)
-\frac{E^2_\eta \left(N_b\right)}{\zeta/a},
\end{equation} 
where $\lambda_\eta\left(N_b,\zeta/a\right)$ is the occupation in a lattice
for a given $N_b$ and $\zeta$, and $E^2_\eta \left(N_b\right)$ is a function 
of the number of particles for each $\eta$.

We calculate $\Lambda_\eta\left(N_b\right)$ and $E^2_\eta \left(N_b\right)$ 
for the three lowest natural orbitals analyzing systems with fillings up to 
401 particles and sizes up to 2000 lattice sites. 
\begin{figure}[h]
\begin{center}
\includegraphics[width=0.47\textwidth,height=0.29\textwidth]
{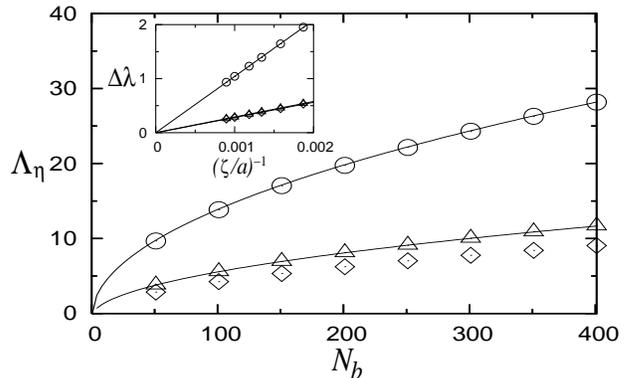}
\end{center} \vspace{-0.7cm}
\caption{Extrapolated values of the first ($\bigcirc$), 
second ($\triangle$), and third ($\Diamond$) natural orbital 
occupations in the continuous system ($\Lambda_0$). 
The lines following the data for the first two natural orbitals 
correspond to the results obtained in Ref.\ \cite{forrester03}. 
The inset shows $\Delta \lambda$ (see text) as a function 
of $(\zeta/a)^{-1}$ also for the first ($\bigcirc$), 
second ($\triangle$), and third ($\Diamond$) natural orbitals in systems 
with 301 HCB's and sizes between 1000 and 2000 lattice sites, 
the straight lines are the result of our fits.}
\label{NatOrbTrapvsN}
\end{figure}
In Fig.\ \ref{NatOrbTrapvsN} we compare our results for 
$\Lambda_\eta\left(N_b\right)$ with the ones presented in 
Ref.\ \cite{forrester03} for the continuous case [see Eqs.\ (91) and (92) there]. 
The agreement is excellent like in the periodic case. 
As an inset we plot $\Delta \lambda=\Lambda_{\eta}\left(N_b\right)
-\lambda_\eta\left(N_b,\zeta/a\right)$ also for the three lowest natural 
orbitals as a function of $(\zeta/a)^{-1}$ confirming Eq.\ (\ref{scalTrap1}). 
(The results obtained for $\Delta \lambda$ for the second and third natural
orbitals are very similar, so in the inset they are one on top of the other.) 
We find that Eq.\ (\ref{scalTrap1}) describes our numerical results up to 
characteristic densities very close to the the one at which the Mott insulator 
appears in the middle of the trap. Hence the dependence of the lowest natural 
orbital occupation on the density in Fig.\ \ref{NatOrbTrapvsNb} can be 
understood as follows.

Comparing Eqs.\ (\ref{scalTrap1}) and (\ref{lambda0HCBGS}), it is possible 
to determine the dependence of $\Lambda_0 \left(N_b\right)$,
\cite{forrester03} and $E^2_0 \left(N_b\right)$ on $N_b$ 
\begin{equation}
\Lambda_{\eta} \left(N_b\right)= b_2\ \sqrt{N_b}, \qquad
E^2_\eta \left(N_b\right)=c_2\ N_b^{3/2},
\end{equation} 
where $b_2$, $c_2$ are parameters that can be determined numerically. 
Both relations above were confirmed by our numerical 
results. Then Eq.\ (\ref{scalTrap1}) may be rewritten as
\begin{equation}
\label{scalTrap2}
\lambda_0\left(N_b,\tilde{\rho}\right)/\sqrt{\zeta/a} = 
\sqrt{\tilde{\rho}} \left( b_2 -c_2\ \tilde{\rho}\right),
\end{equation}
which describes very well the behavior observed in Fig.\ \ref{NatOrbTrapvsNb} 
before the Mott insulator appears in the middle of the system. This is shown 
in Fig.\ \ref{NatOrbTrapvsNb} by a fit to our numerical results for $N=1000$ 
with $b_2=1.44$ and $c_2=0.26$.

\section{Hard-core bosons trapped in other confining potentials}

The results obtained in the previous section for HCB's confined 
in harmonic traps can be generalized for arbitrary 
powers of the confining potential when the appropriate length scale,
\begin{equation}
\zeta=\left(V_{\alpha}/t \right)^{-1/\alpha}, \label{charlengthGS}
\end{equation} 
is considered. [$\alpha$ is the power of the confining potential 
in Eq.\ (\ref{HamHCB})]. In this section, we perform systematically 
the generalization for a power $\alpha=8$ of the confining potential 
in order to show that the statement above holds. 

\begin{figure}[h]
\begin{center}
\includegraphics[width=0.49\textwidth,height=0.43\textwidth]
{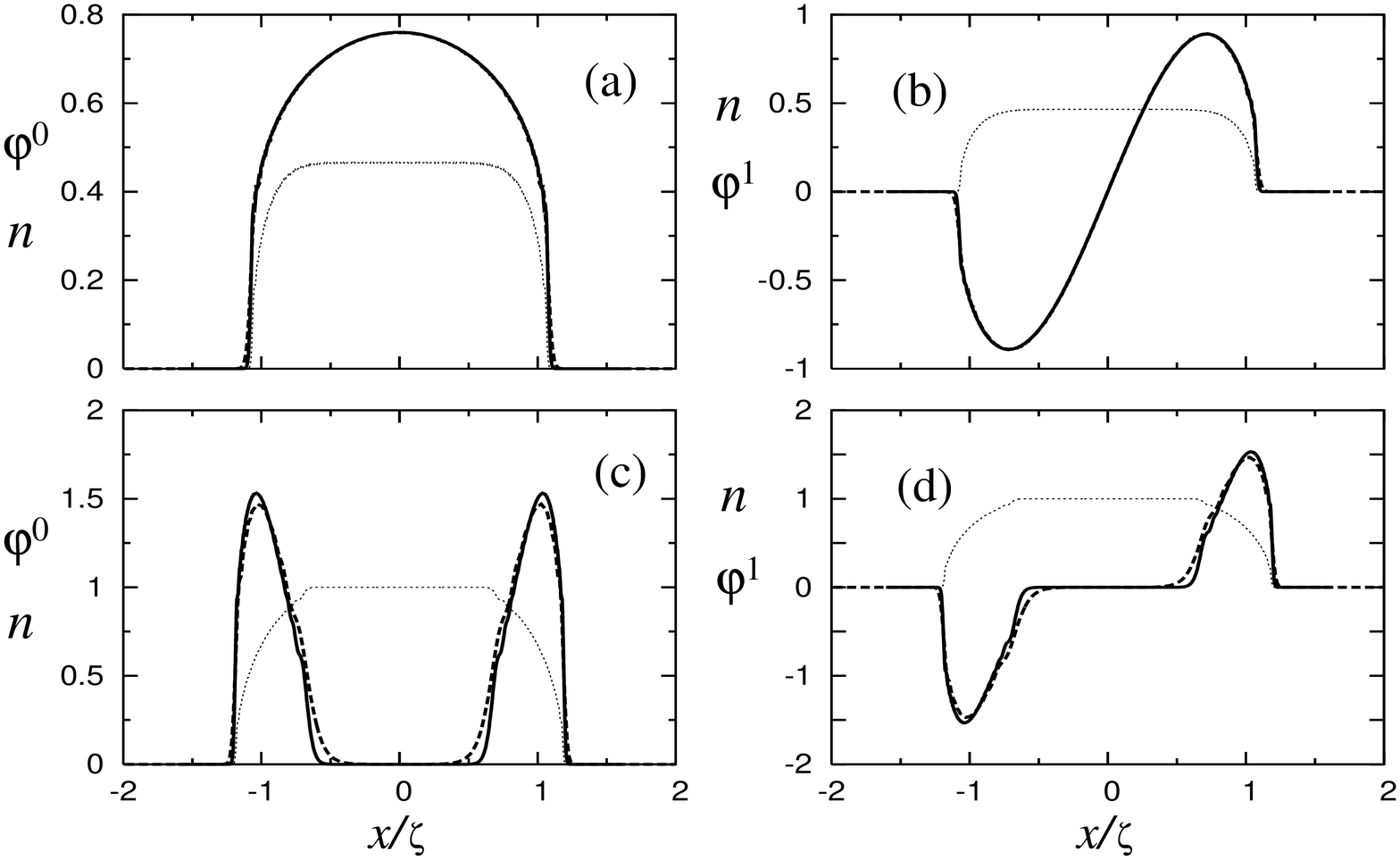}
\end{center} \vspace{-0.7cm}
\caption{Profiles of the two lowest natural orbitals for trapped systems with 
700 lattice sites, $V_8a^8=2.0\times 10^{-19}t$, and occupations of 
201 (a),(b)  and 451 (c),(d) HCB's (continuous line); 
300 lattice sites, $V_8a^8=1.0\times 10^{-15}t$, and occupations of 
69 (a),(b) and 155 (c),(d) HCB's (dashed line). The density profiles have 
been included as thin dotted lines.}
\label{NatOrbWF0700}
\end{figure}
It has been already shown in Ref.\ \cite{rigol03_2} that density 
profiles as a function of the scaled positions in the trap ($x/\zeta$) do not 
change when parameters in the system are changed keeping the characteristic 
density constant. This holds for any power of the confining potential. 
We start this section showing that the scaling of the lowest 
natural orbital, defined by [Eq.\ (\ref{NOSGS})] for the harmonic case 
($\alpha=2$), also holds for $\alpha=8$. In Fig.\ \ref{NatOrbWF0700}, 
we compare the scaled two lowest natural orbitals for systems with different 
filling but the same characteristic density. $\varphi$ was defined like in 
Eq.\ (\ref{NOSGS}) with $\zeta=\left(V_{8}/t \right)^{-1/8}$, as 
corresponds to $\alpha=8$. (Density profiles were included as thin dotted 
lines.) The plots show that indeed the scaling defined holds.

\subsection{Off-diagonal correlations and the lowest natural orbital 
occupation}

\begin{figure}[h]
\begin{center}
\includegraphics[width=0.47\textwidth,height=0.29\textwidth]
{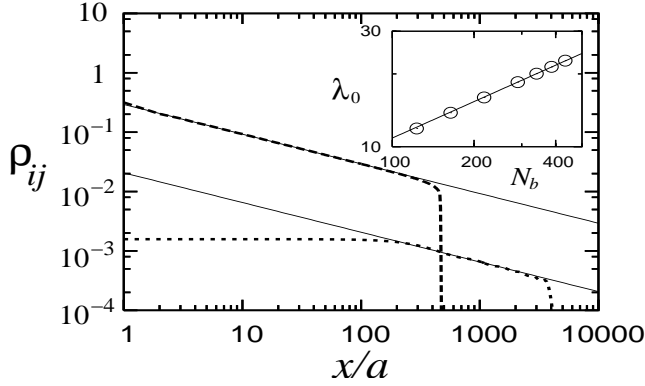}
\end{center} \vspace{-0.7cm}
\caption{HCB one-particle density matrix vs $x/a$ ($x = \mid x_i -x_j \mid$) 
for systems with $N_b=400$, $\tilde{\rho}=0.92$, $n_i=0.46$ (dashed line), 
and for a very dilute system with $N_b=11$, $\tilde{\rho}=7.56\times 10^{-4}$,
$n_i=1.6\times 10^{-3}$ (dotted line). The point $i$ was chosen in the middle of 
the trap and only $j$ was changed. The abrupt reduction of $\rho_{ij}$ occurs 
for $n_j\rightarrow 0$. Thin continuous lines correspond to power laws 
$\sqrt{x/a}$. The inset shows $\lambda_0$ vs $N_b$ for systems with 
$\tilde{\rho}=1$ ($\bigcirc$), the straight line exhibits $\sqrt{N_b}$ 
behavior.}
\label{Largextrap8}
\end{figure}

\begin{figure}[h]
\begin{center}
\includegraphics[width=0.47\textwidth,height=0.28\textwidth]
{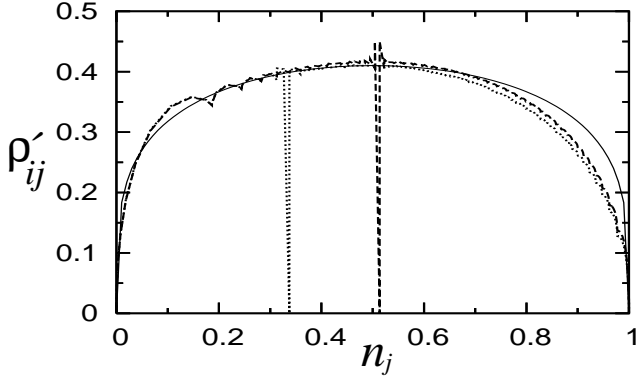}
\end{center} \vspace{-0.7cm}
\caption{HCB one-particle density matrix $\rho'_{ij}$ [Eq.\ (\ref{rhop})] 
vs $n_j$. $\rho'_{ij}$ is measured 
from points $i$ with $n_i=0.51$ (dashed line) and $n_j=0.34$ (dotted line) 
in a trap with $\alpha=8$, $N_b=999$, and $\tilde{\rho}=2.05$ so that a Mott 
insulator is formed in the middle of the system. The thin continuous line 
corresponds to $0.58[n_j(1-n_j)]^{1/4}$, like in Fig.\ \ref{Largexvsrho}.}
\label{Largexvsrho8}
\end{figure}

The increase of the power $\alpha$ of the confining potential leads 
to the formation of more homogeneous density profiles, as shown in 
Fig.\ \ref{NatOrbWF0700}. The fast changes of the density are restricted 
to smaller regions in the borders of the system. In Fig.\ \ref{Largextrap8}, 
it can be seen that also for these systems off-diagonal quasi-long-range 
correlations are present in the one-particle density matrix. Like for the 
harmonic case, they decay as power laws $\rho_{ij}\sim |(x_i-x_j)/a|^{-1/2}$ 
for large distances and away from $n_i,n_j= 0,1$. In the figure, we show 
results for a very dilute case, and for a trap with a density $n_i\sim 0.5$ 
in the middle of the system.
A detailed study shows that like in the harmonic case, 
$\rho_{ij}$ and $|(x_i-x_j)/a|^{-1/2}$ are related by a slowly varying
(away from $n_i,n_j= 0,1$) function of the density. We find this function, 
which we denote as $f_8(n_i,n_j)$, to depart at high densities from the 
$f_2(n_i,n_j)$ [Eq.\ (\ref{rhoN})] behavior found in the harmonic case, 
i.e., the functional dependence of $\rho_{ij}$ on the density {\it is not} 
universal. This can be seen in Fig.\ \ref{Largexvsrho8} where we plot 
$\rho'_{ij}$ [Eq.\ (\ref{rhop})] vs $n_j$ for two different values of $n_i$, 
and compare the results with $0.58[n_j(1-n_j)]^{1/4}$. The differences 
between the plots start to be evident for large densities.

Considering the results above for the power-law decay of the one-particle 
density matrix and for the scaling of the lowest natural orbitals, it is 
possible to generalize for arbitrary powers of the confining potential the 
result obtained in the previous section for the occupation 
of the lowest natural orbital in the harmonic case [Eq.\ (\ref{lambda0HCBGS})].
The differences between $f_8(n_i,n_j)$, $f_2(n_i,n_j)$, and eventually an 
arbitrary $f_\alpha(n_i,n_j)$ do not change the final result in 
Eq.\ (\ref{lambda0HCBGS}) since density profiles as a function of the 
scaled positions $x/\zeta$ are unchanged when the characteristic density 
is kept constant. Then, independently of the power $\alpha$ of the trapping 
potential in Eq.\ (\ref{HamHCB}), the occupation of the lowest natural 
orbital is determined in the thermodynamic limit as
\begin{equation}
\label{lambda0alpha}
\lambda^\alpha_0 =C_\alpha(\tilde{\rho}) \sqrt{N_b}
=D_\alpha(\tilde{\rho}) \sqrt{\zeta/a} \ ,
\end{equation} 
where $C_\alpha(\tilde{\rho})$ and $D_\alpha(\tilde{\rho})$ 
are functions of $\tilde{\rho}$ for a given power of the confining 
potential $\alpha$. The thermodynamic limit is defined keeping the 
characteristic density $\tilde{\rho}=N_b/\zeta$ constant with $\zeta$ 
given by Eq.\ (\ref{charlengthGS}). As for the periodic and harmonic 
cases, the scaling in the thermodynamic limit is valid up to 
a good approximation for finite-size systems, as is shown in the 
inset in Fig.\ \ref{Largextrap8} for $\alpha=8$ and $N_b>100$.

\begin{figure}[h]
\begin{center}
\includegraphics[width=0.45\textwidth,height=0.27\textwidth]
{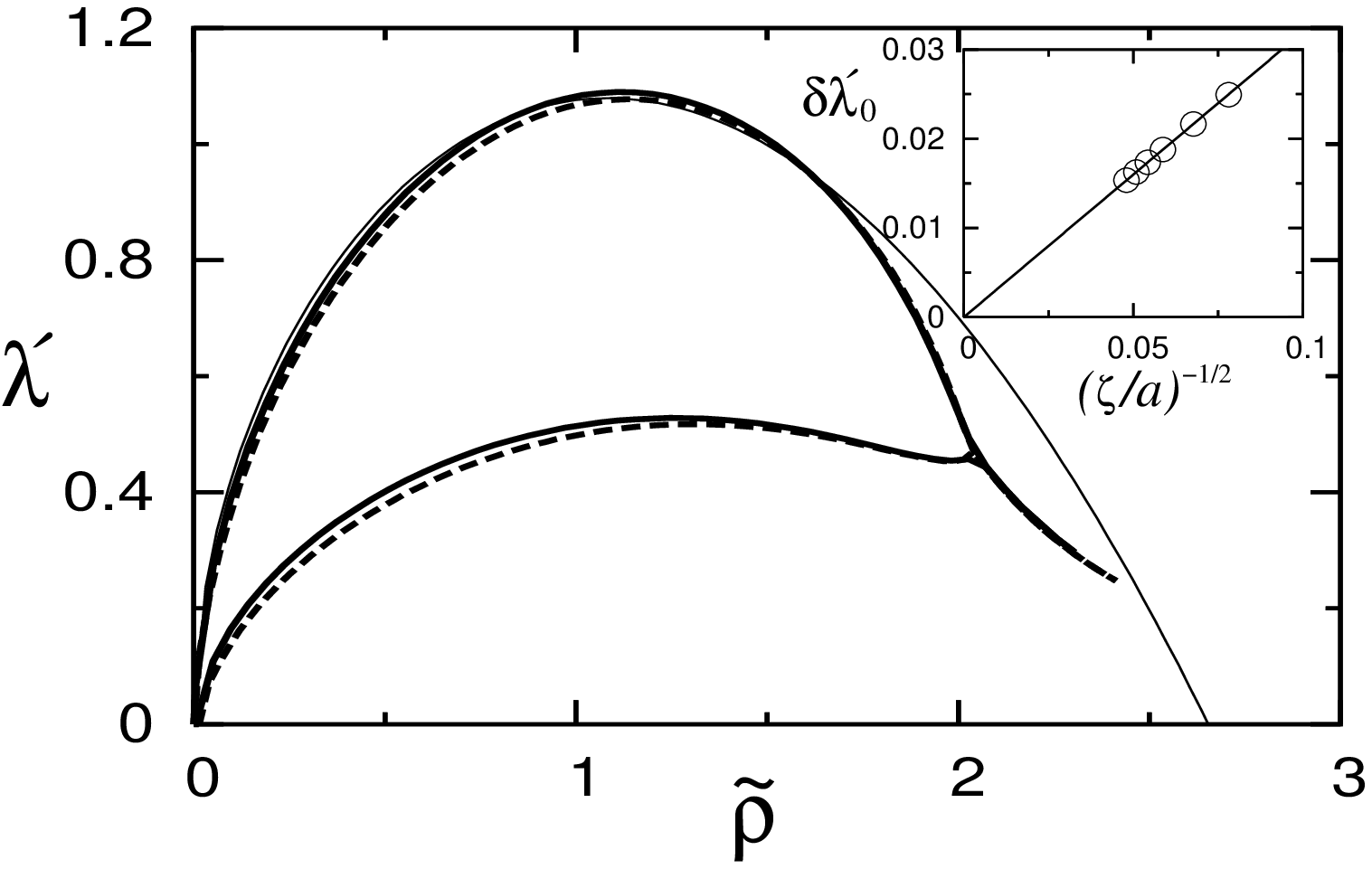}
\end{center} \vspace{-0.7cm}
\caption{Normalized occupation of the two lowest natural orbitals (see text) 
vs $\tilde{\rho}$. The systems analyzed have $N=700$, 
$V_8a^8=2\times 10^{-19}t$, and occupations up to 501 particles 
(thick continuous line), and $N=300$, $V_8a^8=1.0\times 10^{-15}t$, 
and occupations up to 181 particles (dashed line). The thin continuous line 
shows a fit of Eq.\ (\ref{generlambda0}), with $b_8=1.36$ and $c_8=0.29$, 
to our numerical results for $N=700$ ($\gamma=8/5$ for $\alpha=8$). 
The inset displays $\delta \lambda'_0$ vs $(\zeta/a)^{-1/2}$ (see text) 
for $\tilde{\rho}=1.0$ ($\bigcirc$), the continuous line shows the result of 
our fits.}
\label{NatOrbTrap8vsNb}
\end{figure}
A global picture of the occupation of the two lowest natural orbitals in the 
case $\alpha=8$ is shown in Fig.\ \ref{NatOrbTrap8vsNb}(a). There we plot 
$\lambda'=\lambda/\sqrt{\zeta/a}$ vs $\tilde{\rho}$, 
like in the harmonic case (Fig.\ \ref{NatOrbTrapvsNb}), and compare
systems with different curvatures of the confining potential. 
The Mott insulator appears in the middle of the trap when the degeneracy 
of the natural orbitals sets in. The comparison between the plots shows 
that also for $\alpha=8$ and the system sizes chosen, finite-size corrections 
are small. In the following we check whether they have the same 
functional form as for the homogeneous and harmonically trapped cases. 
The inset in Fig.\ \ref{NatOrbTrap8vsNb} shows that this is indeed the case. 
We assume that $\lambda_0/\sqrt{\zeta/a}= 
C_8(\tilde{\rho})-D_8(\tilde{\rho})/\sqrt{\zeta/a}$, 
like for the harmonic trap [Eq.\ (\ref{finsizetrap})], and 
calculate $C_8(\tilde{\rho})$ and $D_8(\tilde{\rho})$. 
In the inset in Fig.\ \ref{NatOrbTrap8vsNb}, we plot 
$\delta\lambda'_0=C_8(\tilde{\rho})-\lambda_0/\sqrt{\zeta/a}$ 
vs $(\zeta/a)^{-1/2}$ for $\tilde{\rho}=1$, the linear behavior observed 
confirms our assumption.

\subsection{Low-density limit in the lattice}

\begin{figure}[h]
\begin{center}
\includegraphics[width=0.48\textwidth,height=0.29\textwidth]
{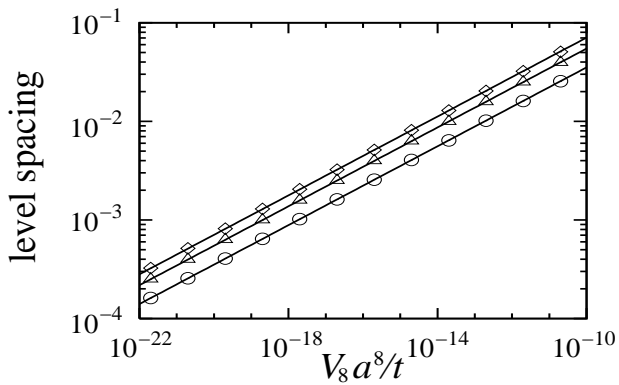}
\end{center} \vspace{-0.7cm}
\caption{Level spacing (in units of $t$) vs $V_8a^8/t$. 
($\bigcirc$) between first and second level, 
($\triangle$) between second and third level, 
and ($\Diamond$) between third and fourth level. 
The lines following the data correspond to power laws
$(V_8a^8/t)^{0.2}$.}
\label{levelspacing}
\end{figure}

In the following, we analyze the corrections introduced by the lattice 
to the occupation of the lowest natural orbitals in the 
confined continuous case. For a trap with power $\alpha=8$ of the 
confining potential, it can be proven numerically 
(Fig.\ \ref{levelspacing}) that in the low-energy region the ratio 
between the level spacing and the bandwidth reduces proportionally 
to $(V_8a^8/t)^{1/5}$, i.e., proportionally to $(\zeta/a)^{-8/5}$.
As for the periodic and harmonically trapped systems, we find that 
the lowest natural orbital occupations approach their value in the 
continuous system $\Lambda_{\eta}\left(N_b\right)$ following the same 
behavior,
\begin{equation}
\label{scalTrap8a}
\lambda_\eta\left(N_b,\zeta/a\right) = \Lambda_{\eta}\left(N_b\right)
-E^8_\eta \left(N_b\right)(\zeta/a)^{-8/5}.
\end{equation} 

\begin{widetext}

\begin{figure}[h]
\begin{center}
\includegraphics[width=0.81\textwidth,height=0.33\textwidth]
{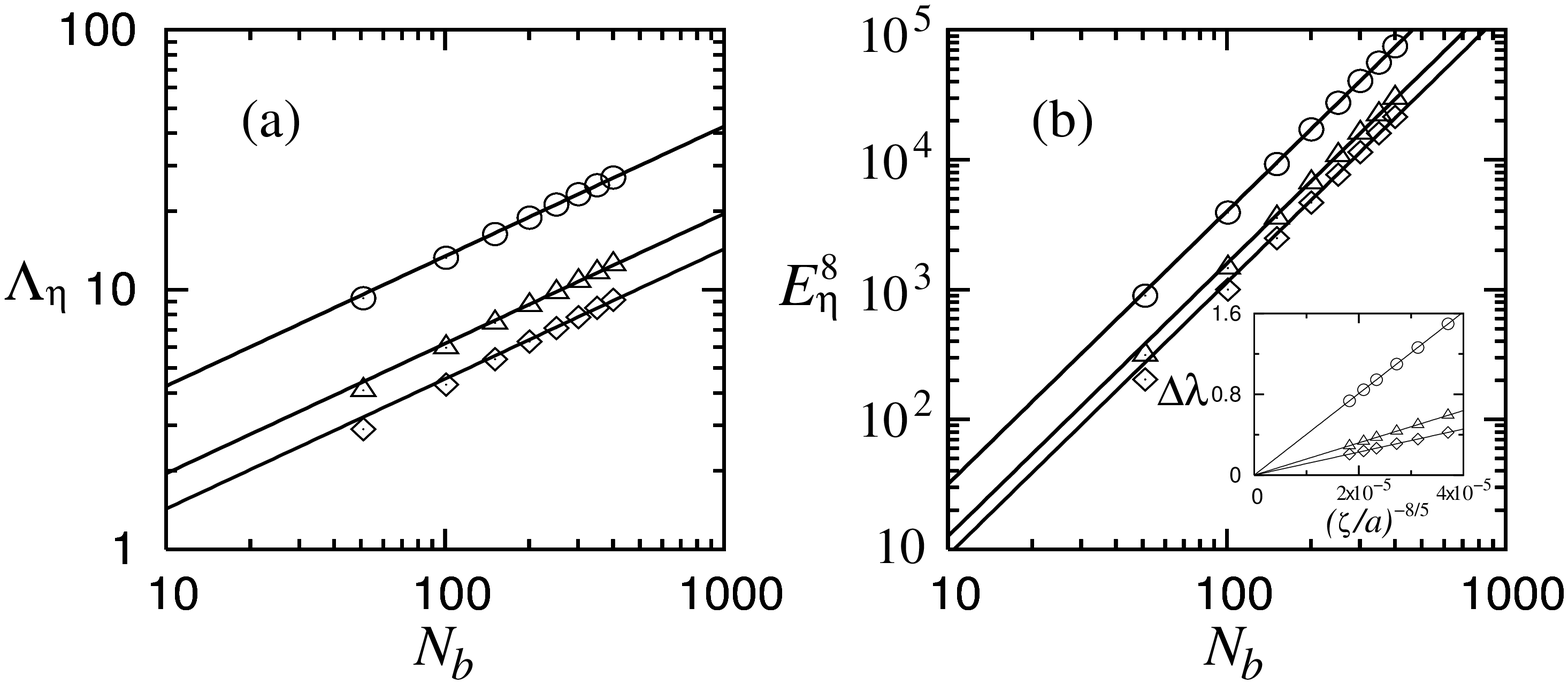}
\end{center} \vspace{-0.7cm}
\caption{Values obtained for $\Lambda_{\eta}\left(N_b\right)$ (a)
and $E^8_\eta \left(N_b\right)$ (b) vs $N_b$, 
for the first ($\bigcirc$), second ($\triangle$), 
and third ($\Diamond$) natural orbitals. The lines following the data 
correspond to power laws $\sqrt{N_b}$ (a) and 
$N_b^{21/10}$ (b). 
The inset in (b) shows $\Delta \lambda$ (see text) as a function 
of $(\zeta/a)^{-8/5}$ also for the first ($\bigcirc$), 
second ($\triangle$), and third ($\Diamond$) natural orbitals in systems 
with 301 HCB's and sizes between 1000 and 2000 lattice sites.}
\label{NatOrbTrap8vsN}
\end{figure}

\end{widetext}

We calculate $\Lambda_\eta\left(N_b\right)$ and $E^8_\eta \left(N_b\right)$ 
for the three lowest natural orbitals analyzing systems with fillings up to 
401 particles and sizes up to 2000 lattice sites. The results obtained are 
shown in Figs.\ \ref{NatOrbTrap8vsN}(a) and \ref{NatOrbTrap8vsN}(b), 
respectively. In Fig.\ \ref{NatOrbTrap8vsN} we compare our results 
for $\Lambda_\eta\left(N_b\right)$ and $E^8_\eta \left(N_b\right)$ 
with the power laws expected from comparing 
Eq.\ (\ref{scalTrap8a}) and Eq.\ (\ref{lambda0alpha}), i.e.,
\begin{equation}
\Lambda_\eta\left(N_b\right)=b_8\ \sqrt{N_b}, \qquad 
E_8\left(N_b\right)=c_8\ N_b^{21/10}
\end{equation}

The plots show that for large occupations, basically larger that 100 HCB's, 
the expected power laws are observed. In addition, we plot as an inset 
in Fig.\ \ref{NatOrbTrap8vsN}(b) 
$\Delta \lambda=\Lambda_{\eta}\left(N_b\right)
-\lambda_\eta\left(N_b,\zeta/a\right)$ also for the three lowest 
natural orbitals as a function of $(\zeta/a)^{-8/5}$. These plots show that 
the scaling behavior proposed in Eq.\ (\ref{scalTrap8a}) is correct. We also 
find that like in the harmonic trap, Eq.\ (\ref{scalTrap8a}) follows the 
numerical results up to fillings very close to the one at which the 
Mott insulator appears in the middle of the system. Equation (\ref{scalTrap8a}) 
can be rewritten for the lowest natural orbital as
\begin{equation}
\lambda_0\left(N_b,\zeta/a\right) = \sqrt{N_b}
\left( b_8-c_8\ \tilde{\rho}^{\gamma}\right) 
\label{generlambda0}
\end{equation}
with $\gamma=8/5$. Similar expressions were obtained for the 
periodic (changing $\tilde{\rho}$ by $\rho$) [Eq.\ (\ref{scalHom2})] and 
harmonically trapped [Eq.\ (\ref{scalTrap2})] 
systems, with $\gamma=2$ and $\gamma=1$, respectively. 
As is shown in Fig.\ \ref{NatOrbTrap8vsNb}, Eq.\ (\ref{generlambda0}) 
(with $\gamma=8/5$) describes very well our numerical results below 
$\tilde{\rho}=2$ for the following fitting parameters: 
$b_8=1.36$ and $c_8=0.29$. 

\section{Conclusions}

We have studied in detail ground-state properties of hard-core bosons 
confined on 1D lattices. The results obtained for the off-diagonal 
behavior of the one-particle density matrix show that it decays 
with the same power law known from periodic systems \cite{mccoy68}, 
independently of the power of the confining potential. In contrast, 
we find that the small dependence of the one-particle density matrix 
on the density is not universal. In the harmonically trapped case 
we were able to find a function that fits very well this density 
dependence [Eq.\ (\ref{f2})]. The results we have obtained for HCB's 
are equally valid for the 1D $XY$ model with a space-varying transverse 
field. 

We have shown how the occupation of the lowest natural orbital is set by 
the large distance behavior of the one-particle density matrix. Even in 
the cases where a region with $n_i=1$ builds up in the middle of the system, 
we find that this quantity scales proportionally to the square root of the 
number of particles (at constant characteristic density), and independently 
of the power of the confining potential. The functional form of the constant 
of proportionality, as a function of the characteristic density, was 
determined for fillings below the one at which the Mott insulating region 
appears in the density profiles. We have also obtained the first correction 
to the square-root behavior of the lowest natural orbital occupation due to 
finite-size effects. We find that it is also universal independent of the 
power of the confining potential. Although not discussed here, we should 
mention that the natural orbital occupations $\lambda_\eta$ also display 
another universal behavior for large values of $\eta$ at low densities 
$\lambda_\eta\sim \eta^{-4}$, a behavior that is shared by the 
momentum distribution function for large momenta $n_k\sim |k|^{-4}$ 
\cite{rigol04_1}.

Finally, we studied systematically the low-density limit in a lattice. We have 
shown how the results for the occupation of the lowest natural orbitals in 
continuous systems can be obtained as an extrapolation in a lattice. The only 
input knowledge ones needs to do this extrapolation is the behavior of the 
ratio between the level spacing and the bandwidth for the lowest energy levels 
as a function of the strength of the confining potential.

\begin{acknowledgments}

We would like to thank HLR-Stuttgart (Project DynMet) for allocation of 
computer time and SFB 382 for financial support. We are grateful to 
S. Wessel for useful discussions.

\end{acknowledgments}

\end{document}